\documentclass[journal]{IEEEtran}

\usepackage{booktabs}
\usepackage{tabularx}

\usepackage{amssymb}
\usepackage{amsbsy}
\usepackage{amsmath}
\usepackage{bm}
\usepackage{verbatim}
\usepackage{cite}
\usepackage{mathrsfs}
\usepackage{amsfonts}
\usepackage{graphicx}
\usepackage[tight,footnotesize]{subfigure}
\usepackage[10pt]{moresize}
\usepackage{array}
\usepackage{color}
\usepackage{epsfig}
\usepackage{stfloats}
\usepackage{balance}

\usepackage[noend]{algpseudocode}

\usepackage{algorithmicx,algorithm}

\usepackage{cite}

\usepackage{setspace}
\usepackage{cases}

\usepackage{graphicx}
\usepackage{epstopdf}
\usepackage{multirow}
\usepackage{extarrows}

\newcommand{\subparagraph}{}
\usepackage{titlesec}
\titlespacing{\section}{0pt}{1.0ex plus 0.2ex minus 0.2ex}{0.8ex plus 0.1ex}
\titlespacing{\subsection}{0pt}{0.8ex plus 0.2ex minus 0.2ex}{0.6ex plus 0.1ex}
\titlespacing{\subsubsection}{0pt}{0.6ex plus 0.1ex minus 0.1ex}{0.5ex plus 0.1ex}
\usepackage{amsmath} 
\allowdisplaybreaks[4]

\ifCLASSINFOpdf

\else

\fi

\begin{document}

\title{\Huge Joint Antenna Positioning and Beamforming for Movable Antenna Array Aided Ground Station in Low-Earth Orbit Satellite Communication}

\author{ Jinming~Wang,~\IEEEmembership{Graduate~Student~Member,~IEEE,} Lipeng~Zhu,~\IEEEmembership{Member,~IEEE,}  Shuai~Han,~\IEEEmembership{Senior~Member,~IEEE,} He~Sun,~\IEEEmembership{Member,~IEEE,} and Rui~Zhang,~\IEEEmembership{Fellow,~IEEE}.
	\thanks{
		Jinming~Wang and Shuai~Han are with the School of Electronics and Information Engineering, Harbin Institute of Technology, China (e-mails: jinm\_wang@163.com and hanshuai@hit.edu.cn). 
		
		Lipeng Zhu, He Sun, and Rui Zhang are with the Department of Electrical and Computer Engineering, National University of Singapore, Singapore 117583 (e-mails: zhulp@nus.edu.sg, sunele@nus.edu.sg, elezhang@nus.edu.sg). }
}

\maketitle
\pagestyle{empty}
\thispagestyle{empty}

\begin{abstract}
This paper proposes a new architecture for the low-earth orbit (LEO) satellite ground station aided by movable antenna (MA) array. Unlike conventional fixed-position antenna (FPA), the MA array can flexibly adjust antenna positions to reconfigure array geometry, for more effectively mitigating interference and improving communication performance in ultra-dense LEO satellite networks. To reduce movement overhead, we configure antenna positions at the antenna initialization stage, which remain unchanged during the whole communication period of the ground station. To this end, an optimization problem is formulated to maximize the average achievable rate of the ground station by jointly optimizing its antenna position vector (APV) and time-varying beamforming weights, i.e., antenna weight vectors (AWVs). To solve the resulting non-convex optimization problem, we adopt the Lagrangian dual transformation and quadratic transformation to reformulate the objective function into a more tractable form. Then, we develop an efficient block coordinate descent-based iterative algorithm that alternately optimizes the APV and AWVs until convergence is reached. Simulation results demonstrate that our proposed MA scheme significantly outperforms traditional FPA by increasing the achievable rate at ground stations under various system setups, thus providing an efficient solution for interference mitigation in future ultra-dense LEO satellite communication networks.
\end{abstract}


\IEEEpeerreviewmaketitle

\section{Introduction}
With the growing demand for seamless global connectivity and ubiquitous access to broadband services, satellite communication systems have gained significant interest as a key enabler for future sixth-generation (6G) mobile communication networks \cite{10045716,10679201,10032237,10899883,10693273}. Compared with traditional terrestrial network infrastructures, low-earth orbit (LEO) satellite systems provide a promising solution for reducing connectivity inequality, especially in remote, rural, or maritime areas where fiber deployment is costly or impractical. Owing to their low altitude, typically ranging from $500$ to $2,000$ km above Earth, LEO satellites offer several advantages over traditional geostationary-earth orbit (GEO) satellites, including reduced propagation delay, lower propagation path loss, and improved capacity for low-latency and high-throughput services \cite{9982369,10436111,10535997}. These characteristics are particularly attractive for delay-sensitive and mission-critical applications. Consequently, LEO satellite networks are being increasingly considered for a wide range of emerging use cases, such as global navigation and positioning, environmental monitoring, remote sensing, disaster recovery, autonomous vehicle communication, and large-scale Internet of things (IoT) deployments \cite{10492466,10736568,10749983}. Motivated by these benefits, several commercial entities, such as Starlink (SpaceX), OneWeb, and Amazon’s Project Kuiper, have initiated large-scale deployments of LEO constellations, involving thousands of satellites in various orbital shells. These initiatives aim to build a global satellite internet infrastructure capable of supporting high-throughput communication services, resilient transmission, and massive connectivity worldwide \cite{9755278,9119123,10286330}. \par
Despite the promising outlook of LEO satellite networks, their unique operational characteristics also introduce several critical challenges that hinder the realization of reliable and high-throughput communications. One critical issue stems from the long propagation path between the satellite and the ground station, which leads to substantial path loss, especially at high carrier frequencies \cite{10794640,10589577,10679605}. To mitigate this, researchers have explored various enhancement techniques on both the satellite and the ground station sides, such as beamforming \cite{9495105} and large-aperture antennas \cite{10355084}. Another fundamental challenge arises from the dense deployment of LEO constellations \cite{9905747,10242022,10480330}. With thousands of satellites simultaneously orbiting the Earth in multiple orbital planes, ground stations are often covered by multiple satellites within their visibility region \cite{10798457}. This results in severe inter-satellite interference, which can significantly degrade the signal reception quality at the ground station and limit the overall system capacity. Consequently, efficient interference suppression techniques are essential for enabling scalable and high-throughput LEO satellite communication networks. \par
To alleviate the impact of inter-satellite interference, a variety of interference suppression techniques have been investigated, with array beamforming emerging as one of the most effective solutions. At the satellite side, exclusion zones \cite{10769081}, beamforming design \cite{10054459}, and resource allocation \cite{10829623} have been proposed to reduce the interference at the ground station. At the ground station side, effective beamforming methods \cite{10507224} enable spatial filtering by adjusting the antenna weight vector (AWV), i.e., beamforming weights, to enhance desired signals while suppressing interference from non-serving satellites. However, these methods typically assume a fixed-position antenna (FPA) geometry at the ground station, limiting the spatial degrees of freedom available for beam pattern design. More importantly, compared to geostationary satellites, LEO satellites move rapidly, resulting in highly dynamic interference patterns that vary significantly over short time scales. As a result, existing FPA-based interference suppression techniques may become ineffective in dense LEO constellations, where adaptive techniques are crucial to maintain robust satellite communication performance. \par
To address the limitations imposed by legacy FPA architectures, this paper investigates the use of movable antenna (MA) arrays at the ground station to enhance interference supression in LEO satellite comuunication networks. Unlike conventional FPA arrays with fixed geometry, MA arrays allow for each antenna element/subarray at the ground station to reposition locally within a confined region \cite{10286328,10945745,10906511}, which is also known as fluid antenna system (FAS) \cite{additionone}. This movement capability introduces a new spatial control mechanism, i.e., the antenna position vector (APV), which can be jointly optimized with the AWV to better adapt to dynamic interference and signal directions. The superiority of MA arrays over traditional FPA architectures in terms of beamforming adaptability has been corroborated by prior research efforts \cite{10278220,10382559}. In particular, \cite{10278220} demonstrates that MA arrays can simultaneously achieve high array gain toward desired signals and form nulls in the directions of multiple interferers by exploiting the reconfigurable geometry. Additionally, \cite{10382559} shows that by iteratively adjusting both the APV and AWV, MA arrays exhibit substantial performance gains in multi-beamforming communication scenarios compared with fixed-geometry FPA arrays. These capabilities make MA a promising solution for interference suppression in various communication environments such as LEO satellite networks. \par
In recent years, MA-assisted communication has attracted growing interest in terrestrial wireless systems due to its ability to improve communication quality without the need for deploying additional antennas. Studies based on field-response and spatial correlation channel models have demonstrated that allowing antenna elements to move locally at the transmitter or receiver can significantly enhance the received signal strength in both narrowband and wideband scenarios \cite{10318061,additionone,10508218,10709885}. MA-enabled MIMO systems have been investigated under both instantaneous and statistical channel state information (CSI) conditions, revealing that optimizing antenna element positions yields considerable gains in channel capacity \cite{10243545,10437006}. In the context of multiuser systems, MA arrays have shown their effectiveness in mitigating inter-user interference and enhancing spatial multiplexing by dynamically adapting the array structure \cite{10354003,10741192,10437926,10909572,10388242,10416896,10458417,10477314,11007274}. Additionally, the MA paradigm has proven valuable for improving physical-layer security through adaptive positioning to suppress eavesdropping and enhance secrecy rates \cite{10416363,10447471,10684758}. For aerial platforms, \cite{10654366} explores the joint optimization of MA placement and UAV trajectory, unlocking new degrees of freedom for enhancing UAV-ground communications. Meanwhile, efficient channel estimation techniques for MA systems have been proposed in \cite{10236898,10497534} based on compressed sensing frameworks that reconstruct the field-response characteristics between spatial regions. To further push the boundary, a six-dimensional MA (6DMA) architecture was introduced in \cite{10752873,10848372,10883029}, enabling simultaneous optimization of 3D placement and rotation of the antenna surfaces to match user spatial distributions. Similar design principles have also been extended to reconfigurable intelligent surface
(RIS)/intelligent reflecting surface (IRS) \cite{10806757}, where the position or orientation of RIS/IRS elements is adjusted for improved propagation control \cite{9722711,10430366}. Furthermore, recent efforts have even explored a new flexible antenna architecture, namely pinching antenna, where the antenna positions can be moved on a dielectric waveguide in a large scale to adapt to varying user distributions and wireless communication requirements \cite{10945421,liu2025pinchingantennasystemspassarchitecture,papanikolaou2025resolvingdoublenearfarproblem}. \par
Despite the recent advancements in MA-enabled communications, the vast majority of existing studies have been confined to terrestrial wireless systems \cite{10278220,10382559,10318061,additionone,10508218,10709885,10243545,10437006,10354003,10741192,10437926,10909572,10388242,10416896,10458417,10477314,10416363,10447471,10684758,10654366,10236898,10497534,10752873,10848372,10806757,9722711,10430366,10883029,10945421,liu2025pinchingantennasystemspassarchitecture,papanikolaou2025resolvingdoublenearfarproblem,11007274}, where the deployment environments and mobility characteristics differ significantly from those in satellite communications. Furthermore, a recent study \cite{10806489} has explored the integration of MA arrays in the context of satellite communications. Specifically, it investigates the deployment of MA arrays on LEO satellites to enhance downlink beam coverage and suppress signal leakage toward interference-prone areas. Then, a time-dependent joint optimization of the satellite-mounted APV and AWV is formulated to minimize interference leakage while satisfying beam coverage requirement and antenna moving constraints. Besides, the authors in \cite{10804611,10978720} also investigated the use of MA arrays at the satellite, where the MA positions were optimized to improve the satellite communication system's performance. Although promising in principle, the satellite-side MA architecture faces several practical limitations. For example, the inclusion of movable hardware on satellites increases system complexity and is subject to strict payload constraints in terms of power, size, and mechanical reliability. In addition, the dynamic adjustment of satellite-mounted MAs would require sophisticated control mechanisms and continuous recalibration under fast orbital dynamics, which may compromise stability and reliability. Moreover, once deployed in orbit, the possibility of hardware maintenance or replacement is extremely limited. In comparison, the ground-station MA deployment turns to be a more practical and cost-effective solution in LEO communication systems. \par
In contrast to satellite-side MA architectures, this paper introduces an alternative scheme where the MA array is deployed at the ground station. This approach offers several compelling advantages from both practical and performance perspectives. Ground stations are inherently less constrained by size, weight, and power limitations as compared to LEO satellites, making it more feasible to integrate movable hardware modules without compromising system reliability or increasing satellite payload complexity. Additionally, by shifting the optimization to the ground side, the antenna position can be dynamically tailored to the local signal environment of each individual ground station. This allows for more flexible, user-specific interference suppression and signal enhancement, which is particularly beneficial in large-scale and interference-prone LEO constellations. To the best of our knowledge, this paper is the first work that applies the MA concept at the ground station for LEO satellite communications. In this paper, we jointly optimize the APV and AWVs of the ground station under practical movement and spacing constraints, fully leveraging spatial degrees of freedom to enhance the average achievable rate in interference-limited satellite communications. The main contributions of this paper are summarized as follows:
\begin{itemize}
	\item We propose a novel MA-assisted LEO satellite ground station architecture, in which each antenna element is capable of locally adjusting its position within a predefined region. To reduce the movement overhead, antenna positions are configured once during the initialization phase and remain fixed throughout the subsequent communication period. In addition, a comprehensive system model is established to characterize the dynamic topology of the LEO constellation, taking into account earth rotation, satellite motion, and the directional antenna gain patterns of satellites.
	
	\item Based on the proposed system model, we formulate a joint optimization problem that aims to maximize the average achievable rate of the ground station over a typical communication interval. This is achieved by jointly optimizing the quasi-static APV and the time-varying AWVs corresponding to different time scales. To ensure practical feasibility, several physical constraints are imposed, including the limited movement range of each antenna element and the minimum allowable distance between adjacent antennas. This formulation captures the spatio-temporal coupling between array geometry and beamforming performance.
	
	\item To tackle the non-convexity of the formulated joint optimization problem, we first apply the Lagrangian dual transformation to decouple the logarithmic rate expression, resulting in a more tractable objective function. Based on the transformed problem, we design an efficient iterative algorithm using the block coordinate descent (BCD) method to alternately update the AWVs and APV at the ground station. The AWV subproblem admits a closed-form solution based on standard optimization techniques, whereas the APV subproblem is tackled using a successive convex approximation (SCA) method.
	
	\item Extensive simulations are conducted to evaluate the performance of the proposed MA scheme under various LEO constellation configurations, with different satellite densities, number of antennas, and transmit power levels. The results consistently show that our approach significantly outperforms traditional FPA schemes in terms of average achievable rate. These findings demonstrate the strong potential of MA arrays in enhancing interference suppression and communication reliability for ultra-dense LEO satellite communication networks.
\end{itemize} \par
The remainder of this paper is organized as follows. Section II describes the system and channel models of the proposed MA-assisted LEO satellite network. Section III formulates the joint optimization problem and presents the iterative solution algorithm. Section IV provides simulation results and performance analysis. Finally, the paper is concluded in Section V. \par
\emph{Notations}:  
$a$, $\mathbf{a}$, $\mathbf{A}$, and $\mathcal{A}$ denote a scalar, a vector, a matrix, and a set, respectively. 
$(\cdot)^\text{T}$ and $(\cdot)^\text{H}$ represent the transpose and conjugate transpose operations, respectively. 
$[\mathbf{a}]_n$ denotes the $n$th entry of vector $\mathbf{a}$.
$\mathbb{C}^{M \times N}$ represents the set of complex matrices or vectors of dimension $M \times N$.  
$|\cdot|$ denotes the amplitude of a complex number or complex vector. $\text{Re}(\cdot)$ represents the real part of a complex number or vector. $\|\cdot\|_2$ denotes the Euclidean norm of a vector. $\mathbf{I}$ represents the identity matrix.  
$\mathcal{CN}(\mu, \sigma^2)$ denotes a circularly symmetric complex Gaussian distribution with mean $\mu$ and variance $\sigma^2$. $\partial f / \partial x$ and $\nabla f$ represent the partial derivative and the gradient of a function $f(x)$, respectively. $\lceil \cdot \rceil$ denotes the ceiling function that rounds up to the nearest integer.
\section{System Model and Problem Formulation}
This section presents the system and channel models, followed by the problem formulation for the proposed MA-assisted LEO satellite internet network. We first establish a three-dimensional (3D) geocentric coordinate system to describe the spatial geometry of the satellite constellation and ground station. Based on this, we model the satellite orbits, wave propagation, and MA array structure. We then derive the achievable rate expression, and formulate a joint optimization problem to maximize the average achievable rate by jointly optimizing the APV and AWVs under practical constraints.
\subsection{System Model}
\begin{figure}[t]
	\begin{center} 
		\includegraphics[width=0.45\textwidth]{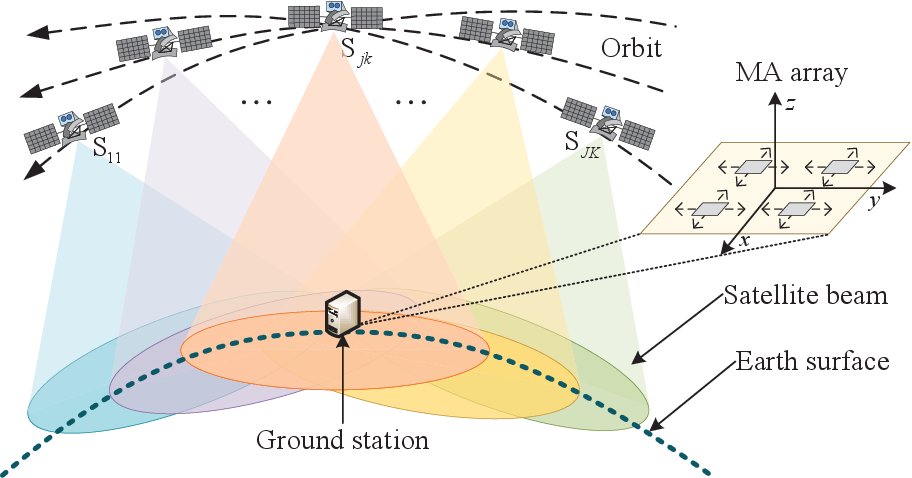}
		\caption{System model of the MA-assisted LEO satellite internet network.}
		\label{fig1}
	\end{center}
\end{figure}

Fig.~\ref{fig1} illustrates the proposed system model of an MA-assisted LEO satellite internet network. In this setup, a ground station located on the earth is equipped with an $N$-element MA array that receives composite signals from multiple satellites in different orbital planes. Without loss of generality, we consider a Walker Delta LEO constellation comprising $J$ orbit planes, each containing $K$ uniformly spaced satellites. Let $\text{S}_{jk}$ denote the $k$th satellite in the $j$th orbit, where $k \in \mathcal{K} \triangleq \{ 1, \cdots, K \}$ and $j \in \mathcal{J} \triangleq \{ 1, \cdots, J \}$. For simplicity, let $\beta$ denote the orbital inclination angle, i.e., the angle between the orbital plane and the equatorial plane. Assuming a perfectly spherical earth with radius $R$, satellite altitude $H$, and earth rotation period $T_\text{E}$, the orbital period $T$ is given by $T = 2 \pi \sqrt{ (R + H)^3/G_\text{e} M_\text{e} }$, where $G_\text{e}$ and $M_\text{e}$ represent the gravitational constant and earth's mass, respectively. \par
\begin{figure}[t]
	\begin{center} 
		\includegraphics[width=0.45\textwidth]{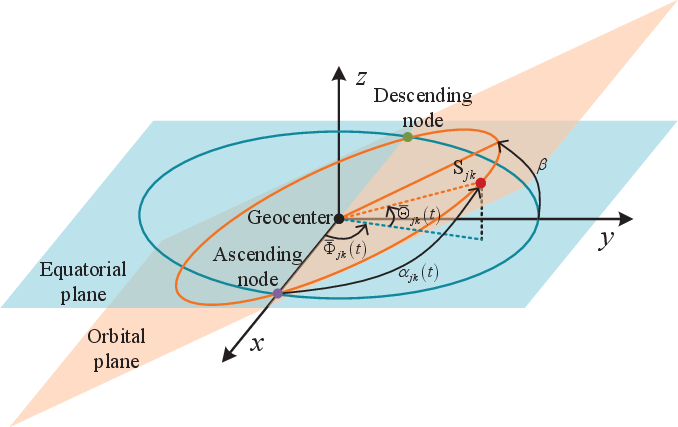}
		\caption{Geocentric coordinate system and the relevant angles.}
		\label{fig2}
	\end{center}
\end{figure}

Based on the system architecture illustrated in \cite{10806489}, we establish a 3D geocentric spherical coordinate system (GSCS), where the equatorial plane serves as the reference plane, as illustrated in Fig.~\ref{fig2}. In this system, the elevation angle \(\Theta \in \left[ -\pi/2, \pi/2 \right]\) is defined as the angle between the reference plane and the direction of a given point, with positive values indicating directions from south to north. Meanwhile, the ascending and descending nodes are defined as the intersection points of the satellite’s orbit with the equatorial plane, where the ascending node corresponds to motion from south to north, and the descending node corresponds to motion from north to south. Define the direction toward the ascending node of the orbit that intersects the equator at the Greenwich meridian as the azimuth reference direction, i.e., the X-axis direction in Fig.~\ref{fig2}. Then, the azimuth angle \(\Phi \in ( -\pi, \pi ]\) denotes the angle between a reference azimuth direction and the projection of the point onto the reference plane, where positive values are measured from west to east. \par
Without loss of generality, we focus on the satellites in the orbital segment traveling from the South Pole to the North Pole, assuming that the satellites in the opposite segment operate in different frequency bands and can adopt a similar optimization strategy. In the established GSCS, let $\alpha_{jk} \left( t \right) = 2 \pi t/T + \alpha_{jk0}$ (in radians) represent the geocentric angle between the ascending node and $\text{S}_{jk}$ at time $t$. $\alpha_{jk0} \in \left[ -\pi/2, \pi/2 \right] $ denotes the initial angle at $t = 0$, which can be given by $\alpha_{jk0} = -\pi/2 + \pi \left( k - 1\right) / \left( K-1 \right) $.
%
%
Accordingly, the position of $\text{S}_{jk}$ in the GSCS is expressed as
\begin{equation}
	\begin{split}
		&\quad\quad\quad\quad\quad\quad \bar{R} = R + H, \\
		&\quad\ \, \bar{\Theta}_{jk} \left( t \right) = \arcsin \left[ \sin \beta \sin \alpha_{jk} \left( t \right) \right], \\
		&\bar{\Phi}_{jk} \left( t \right) =  \arctan \left[ \cos \beta \tan \alpha_{jk} \left( t \right) \right] + \frac{2 \pi j}{ J },
	\end{split}
\end{equation}
where $\bar{R}$ denotes the distance from the satellite to the geocenter, $\bar{\Theta}_{jk}(t) \in \left[ -\pi/2, \pi/2 \right]$ represents the elevation angle measured from the reference plane to the direction of $\text{S}_{jk}$, and $\bar{\Phi}_{jk}(t) \in ( -\pi, \pi ]$ denotes the azimuth angle measured from the reference azimuth direction to the projection of $\text{S}_{jk}$ onto the reference plane. Accordingly, the coordinates of $\text{S}_{jk}$ in the 3D geocentric Cartesian coordinate system (GCCS) are given by
\begin{equation}
	\begin{split}
		\mathbf{R}_{jk} \left( t \right) = &[ \bar{R} \cos\bar{\Theta}_{jk} \left( t \right) \cos \bar{\Phi}_{jk} \left( t \right), \\
		&\quad\quad \bar{R} \cos\bar{\Theta}_{jk} \left( t \right) \sin \bar{\Phi}_{jk} \left( t \right), \bar{R} \sin\bar{\Theta}_{jk} \left( t \right) ]^\text{T}.
	\end{split}
\end{equation}
Moreover, note that satellites within the same orbital plane share identical trajectories, differing only in their relative positions along the orbit. Therefore, analyzing a single representative satellite per orbit is sufficient to capture the system behavior within each plane. Meanwhile, due to earth's rotation, the visible satellite constellation from a given ground station changes periodically over time. Considering the total number of orbits \( J \) and the earth rotation period \( T_\text{E} \), we define a representative observation interval as \( (0, \bar{T}] \), where \( \bar{T} = T_\text{E}/J \) represents the time required for the next orbital plane to reach the same relative position over the ground station. For simplicity, we further assume that \( \bar{T} \) is an integer multiple of the intra-orbit satellite spacing period \( T/K \), which facilitates time slot alignment over the observation period. \par
Then, the position of the ground station is expressed as
%
\begin{equation}
	\begin{split}
		\mathbf{R}_\text{u} \left( t \right) = &[ R \cos \Theta_\text{u} \cos \Phi_\text{u} \left( t \right), \\
		&\quad\quad\quad\quad\quad R \cos \Theta_\text{u} \sin \Phi_\text{u} \left( t \right), R \sin \Theta_\text{u} ]^\text{T},
	\end{split}
\end{equation}
where $\Theta_\text{u}$ and $\Phi_\text{u} \left( t \right) = 2 \pi t /T_\text{E}$ denote the elevation angle and azimuth angle of the ground station, respectively. 
\subsection{Channel Model}
Based on the established system model, the wave vector from the ground station to $\text{S}_{jk}$ in the GCCS can be expressed as $\mathbf{a}_{jk} \left( t \right) = 2 \pi \bar{\mathbf{a}}_{jk} \left(t \right) /\lambda \left\| \bar{\mathbf{a}}_{jk} \left(  t \right) \right\|_2$,
%
%
where $\bar{\mathbf{a}}_{jk}(t) = \mathbf{R}_\text{u}(t) - \mathbf{R}_{jk}(t)$, and $\lambda$ denotes the carrier wavelength. To facilitate channel modeling, we simultaneously establish a ground station-centric Cartesian coordinate system (SCCS). In the SCCS, the $y$-axis is defined as the tangential direction along the latitude line of the ground station, pointing from west to east; the $z$-axis points radially outward from the geocenter; and the $x$-axis is orthogonal to the plane formed by the $y$- and $z$-axes. Given this coordinate setup, the corresponding wave vector expressed in the SCCS can be obtained via the transformation matrix $\mathbf{T}$ as $\tilde{\mathbf{a}}_{jk} \left( t \right) = \mathbf{T}\left( t \right)^\text{T} \mathbf{a}_{jk} \left( t \right)$,
%
%
where
\begin{align}
	\scalebox{0.83}{$
	\mathbf{T}\left( t \right) = \left[ \begin{matrix}
		\sin \Phi_\text{u} \left( t \right) & -\cos \Phi_\text{u} \left( t \right) \sin \Theta_\text{u} & -\cos \Phi_\text{u} \left( t \right) \cos \Theta_\text{u} \\
		-\cos \Phi_\text{u} \left( t \right) & \sin \Phi_\text{u} \left( t \right) \sin \Theta_\text{u} & - \sin \Phi_\text{u} \left( t \right) \cos \Theta_\text{u} \\
		0 & \cos \Theta_\text{u} & - \sin \Theta_\text{u}
	\end{matrix}\right].
$}
\end{align} \par
%
Assuming that each element of the ground station's MA can move within a two-dimensional (2D) region \(\mathcal{A}\) in the $x$-$y$ plane, the position of the $n$th element in the SCCS is denoted by \(\mathbf{c}_n = [x_n, y_n]^\text{T}\), where \(n \in \mathcal{N} \triangleq \{1, \ldots, N\}\). This position is referred to as the APV. Accordingly, the MA array steering vector associated with $\text{S}_{jk}$ is given by
\begin{align}
	\mathbf{s}_{jk} \left(\tilde{\mathbf{a}}_{jk} ( t ), \mathbf{c} \right) = \left[e^{j \tilde{\mathbf{a}}_{jk} ( t )^{\text{T}} \mathbf{c}_n }\right]^\text{T}_{n \in \mathcal{N}},
\end{align}
where \(\mathbf{c} = \left[ \mathbf{c}_1^\text{T}, \cdots, \mathbf{c}_n^\text{T}, \cdots, \mathbf{c}_N^\text{T} \right]^\text{T} \in \mathbb{C}^{2N \times 1} \) represents the collection of \(\mathbf{c}_n\) for all \(n \in \mathcal{N}\). It is important to note that the steering vector fundamentally captures the phase shifts across the MA elements induced by a plane wave characterized by the wave vector \(\tilde{\mathbf{a}}_{jk} ( t )\). In addition, let $\gamma$ and $\rho_0$ denote the path loss exponent and the path gain at a reference distance of 1 meter, respectively. The path gain between $\text{S}_{jk}$ and the ground station is then given by $\rho_{jk} \left( t \right) = \rho_0 \left\| \bar{\mathbf{a}}_{jk} \left( t \right) \right\|_2^{-\gamma}$. \par
Assume that at time $t$, the ground station selects $\text{S}_{jk}$ as the service satellite, while signals from the remaining satellites are treated as co-channel interference. Consequently, the received signal of the ground station at time $t$ can be given by
%
%
\begin{equation}
	\begin{split}
		y \left( \mathbf{c}, \mathbf{w} \left( t\right), t \right) &= \mathbf{w}\left( t \right)^{\text{H}} \mathbf{n} \left( t \right) \\
		&+ \sqrt{P_\text{s}} \sum\limits_{k \in \mathcal{K}_j \left( t \right) \atop j \in \mathcal{J}} \mathbf{w}\left( t \right)^{\text{H}} \mathbf{h}_{jk} \left( \mathbf{c}, t \right) x_{jk} \left( t \right), 
	\end{split}
\end{equation}
where \(P_\text{s}\) denotes the transmit power of the satellites. The set \(\mathcal{K}_j(t) \triangleq \left\lbrace 1, \cdots, k, \cdots, K_j(t) \right\rbrace\) represents the indices of visible satellites in the \(j\)th orbital plane at time \(t\), where \(K_j(t)\) denotes the number of satellites whose signals can be received by the ground station. Specifically, a satellite \(\text{S}_{jk}\) is considered visible if the squared Euclidean distance between the ground station and the satellite satisfies \(\left\| \bar{\mathbf{a}}_{jk}(t) \right\|_2^2 \le (R + H)^2 - R^2\), indicating that the geocentric angle between the satellite and the ground station is within the coverage range of the satellite beam. Besides, the vector \(\mathbf{w}(t) \in \mathbb{C}^{N \times 1}\) denotes the AWV applied at the ground station, while \(\mathbf{n}(t)\) represents independent and identically distributed (i.i.d.) additive white Gaussian noise (AWGN), modeled as \(\mathbf{n}(t) \sim \mathcal{CN}(\mathbf{0}, \sigma^2 \mathbf{I})\). Additionally, \(x_{jk}(t)\) denotes the transmitted data symbol from \(\text{S}_{jk}\). As a result, the corresponding channel vector \(\mathbf{h}_{jk}\) is given by \footnote{During the optimization period, it is assumed that only the line-of-sight (LoS) path is utilized to optimize the MA position. After the optimization, the MA positions remain fixed for the subsequent communication time slots, where CSI can be obtained by conventional pilot-assisted estimation methods tailored to FPA systems. Hence, the antenna position optimization is based on partial CSI information (i.e., the LoS component), which makes this assumption practical in satellite-ground communication systems.}
%
%
\begin{equation}
	\begin{split}
		&\mathbf{h}_{jk} \left( \mathbf{c}, t \right) \\
		&\quad = \sqrt{\rho_{jk} \left( t \right) D\left( \tau_{jk} \left( t \right)  \right) } e^{j \frac{2 \pi}{\lambda} \left\| \bar{\mathbf{a}}_{jk} \left( t \right) \right\|_2}  \mathbf{s}_{jk} \left(\tilde{\mathbf{a}}_{jk} (t), \mathbf{c}\right), 
	\end{split}
\end{equation}
where \(D\left( \tau_{jk}(t) \right)\) represents the directional antenna gain of \(\text{S}_{jk}\). For simplicity and to reflect practical LEO satellite operations, we assume that the satellite beam always points vertically downward. Flexible beam steering at the satellite side may require increasing hardware complexity and power consumption, whereas the ground-station MA provides the required adaptability. \(\tau_{jk}(t)\) denotes the angle between the wave vector \(\bar{\mathbf{a}}_{jk}(t)\) and the opposite direction of the satellite's position vector, i.e., \(-\mathbf{R}_{jk}(t)\). Following the 3GPP technical report TR 38.811 \cite{3GPP38811}, the directional antenna gain pattern is modeled as
%
%
%
\begin{equation}
	\begin{split}
			D&\left( \tau_{jk} \left( t \right)  \right) \\
			& = \begin{cases}
			\quad \quad D_0, \quad  \ \tau_{jk} \left( t \right) = 0, \\
			4 D_0 \left| \frac{J_1(\frac{2 \pi}{\lambda} r \sin (\tau_{jk} \left( t \right)))}{\frac{2 \pi}{\lambda} r \sin (\tau_{jk} \left( t \right))}\right|,\ \ 0 \le \left| \tau_{jk} \left( t \right) \right| \le \frac{\pi}{2},
		\end{cases}
	\end{split}
\end{equation}
where \(J_1(\cdot)\) denotes the Bessel function of the first kind and first order, and \(r\) represents the radius of the antenna’s circular aperture. The maximum antenna gain is given by \(D_0 = \eta (2\pi r / \lambda)^2\), where \(\eta\) denotes the aperture efficiency. \par
Correspondingly, the ground station's signal-to-interference-plus-noise ratio (SINR) at time $t$ is represented as
\begin{align}
	\gamma \left( \mathbf{c}, \mathbf{w} \left( t\right), t \right) = \frac{ \left| \mathbf{h}_{jk} \left( \mathbf{c}, t \right)^\text{H} \mathbf{w} \left( t \right) \right|^2}
	{G_\text{I} + \frac{\sigma^2}{P_\text{s}} \left\|  \mathbf{w} \left( t \right) \right\|^2 },
\end{align}
where $G_\text{I}$ denotes the interference channel gain of other satellites, which is expressed as
\begin{align}
	\scalebox{0.89}{$
	G_\text{I} \triangleq \sum\limits_{l \neq k \atop l \in \mathcal{K}_j \left( t\right) } \left| \mathbf{h}_{jl} \left( \mathbf{c}, t \right)^\text{H} \mathbf{w} \left( t \right) \right|^2 + \sum\limits_{ l \in \mathcal{K}_p \left( t\right) \atop p \neq j, p \in \mathcal{J} } \left| \mathbf{h}_{pl} \left( \mathbf{c}, t \right)^\text{H} \mathbf{w} \left( t \right) \right|^2.
	$}
\end{align}
%
%
Based on the above setup, the achievable rate per unit bandwidth at time $t$ is given by $C \left( \mathbf{c}, \mathbf{w} \left( t\right), t \right) = \log_2 \left( 1 + \gamma \left( \mathbf{c}, \mathbf{w} \left( t\right), t \right) \right)$.
%
%
The ground station's achievable rate is influenced by both the APV and the AWV, in contrast to traditional FPA systems that rely solely on the AWV. Specifically, optimizing the antenna positions in MA arrays enables dynamic reshaping of the steering vectors across different directions. In the following sections, we analyze the interference suppression benefits of MA arrays through the joint optimization of APV and AWV.
\subsection{Problem Formulation}
Note that suppressing interference directly contributes to enhancing the SINR at the ground station. Therefore, for a fixed ground station position, this paper aims to maximize the average achievable rate over the time interval $(0, \bar{T}]$, which is
\begin{align}
	\bar{C} \left( \mathbf{c}, \mathbf{w}(t) \right) = \frac{1}{\bar{T}} \int_0^{\bar{T}} C \left( \mathbf{c}, \mathbf{w}(t), t \right) \mathrm{d}t. 
\end{align}
To address the infinite-dimensional nature introduced by the continuous-time formulation, we discretize the interval \((0,\bar{T}]\) into \(M\) equal-length time slots \(\left[(m-1)\bar{T}/M,\, m\bar{T}/M\right]\), where \(m \in \mathcal{M} \triangleq \{1, \ldots, M\}\). The number of time slots \(M\) is chosen sufficiently large such that the satellite angles \(\bar{\Theta}_{jk}(t)\) and \(\bar{\Phi}_{jk}(t)\) can be considered approximately constant within each slot. We use the midpoint \(t_m = (m - 1/2)\bar{T}/M\) to represent the $m$th slot. Let \(\mathbf{w}[m] \triangleq \mathbf{w}(t_m)\); then, we define
\begin{align}
	\mathbf{w} = \left[ \mathbf{w} \left[1\right]^\text{T}, \cdots, \mathbf{w} \left[m\right]^\text{T}, \cdots, \mathbf{w} \left[M\right]^\text{T} \right]^\text{T} \in \mathbb{C}^{MN \times 1}.
\end{align} \par
Since adjusting the APV in each time slot would result in significant movement overhead and energy consumption, making it challenging to implement, the ground station adopts a common, but optimized MA array geometry across all \(M\) time slots. In other words, the MA array configures its geometry only once during initialization. Accordingly, the average achievable rate over the time interval \((0, \bar{T}]\) is given by
\begin{align}
	\bar{C} \left( \mathbf{c}, \mathbf{w} \right) = \frac{1}{M} \sum\limits_{m \in \mathcal{M}} \log_2 \left( 1 + \gamma_m \right),
\end{align}
where the ground station's SINR $\gamma_m$ is expressed as
\begin{align}
	\frac{ \bar{D}_{jk} \left( t_m \right) \left| \mathbf{s}_{jk} \left(\tilde{\mathbf{a}}_{jk} (t_m), \mathbf{c}\right)^\text{H} \mathbf{w} \left( m \right) \right|^2}
	{ \bar{G}_I + \frac{\sigma^2} {P_\text{s}} \left\|  \mathbf{w} \left( m \right) \right\|^2 },
\end{align}
where \( \bar{D}_{jk} \left( t_m \right) \triangleq D\left( \tau_{jk} \left( t_m \right) \right) \rho_{jk} \left( t_m \right)\), with \(j \in \mathcal{J}\) and \(k \in \mathcal{K}\). The interference channel gain from other satellites is then rewritten as
\begin{equation}
	\begin{split}
		\bar{G}_I \triangleq &\sum\limits_{l \neq k \atop l \in \mathcal{K}_j \left( t_m\right) } \bar{D}_{jl} \left( t_m \right) \left| \mathbf{s}_{jl} \left(\tilde{\mathbf{a}}_{jl} (t_m), \mathbf{c}\right)^\text{H} \mathbf{w} \left( m \right) \right|^2 \\
		&+ \sum\limits_{ l \in \mathcal{K}_p \left( t_m\right) \atop p \neq j, p \in \mathcal{J} }\bar{D}_{pl} \left( t_m \right) \left| \mathbf{s}_{pl} \left(\tilde{\mathbf{a}}_{pl} (t_m), \mathbf{c}\right)^\text{H} \mathbf{w} \left( m \right) \right|^2.
	\end{split}
\end{equation}
Consequently, the optimization problem is given by
\begin{equation}
	\begin{split}
		(\text{P1}) \quad {\mathop{\max}\limits_{\mathbf{c}, \mathbf{w} }} \quad & \sum\limits_{m \in \mathcal{M}} \log_2 \left( 1 + \gamma_m \right) \\ 
		s.t. \quad 
		&\text{C1}: \mathbf{c}_n \in \mathcal{A}, \\
		&\text{C2}: \left\| \mathbf{c}_n - \mathbf{c}_{\tilde{n}} \right\|_2 \ge d_\text{min}, \\
		&\text{C3}: m \in \mathcal{M}, \left\lbrace n, \tilde{n} \right\rbrace \in \mathcal{N}, n \neq \tilde{n},
	\end{split}
\end{equation}
where constraint C1 defines the permissible region within which the antennas can move, while constraint C2 ensures that the inter-antenna spacing is no less than the minimum allowable distance given by \(d_\text{min}\). Notably, the optimization problem (P1) is non-convex and challenging to solve. First, the optimization variables include both the positions and weights of the \(N\) MA elements, resulting in a high-dimensional search space. Second, the coupling between the APV and the AWV across different time slots further complicates the problem. As a result, deriving a globally optimal solution for (P1) is highly challenging.
\section{Optimization Algorithm}
To address the joint optimization of the AWV and APV in the proposed MA-assisted LEO satellite system, this section develops an efficient design framework. We begin by reformulating the original non-convex problem using the Lagrangian dual transformation and quadratic transformation to decouple the complex terms in the objective. Then, by leveraging the BCD method, we alternately update the AWV and APV by solving two subproblems in each iteration. The AWV admits a closed-form solution, while the APV is updated using an SCA approach. Finally, we analyze the convergence and computational complexity of the proposed overall algorithm.
\subsection{Problem Transformation}
To handle the summation of logarithmic terms in the original objective, we first apply the Lagrangian dual transformation, which introduces a set of auxiliary variables \(\alpha_m\), one for each time slot \(m \in \mathcal{M}\). This transformation allows the logarithmic function to be expressed in a more tractable form:
%
%
%
\begin{equation}
	\begin{split}
		\sum\limits_{m \in \mathcal{M}} \log_2 \left( 1 + \gamma_m \right) = \mathop{\max}\limits_{\alpha_m \ge 0} &\ \sum\limits_{m \in \mathcal{M}} \big[ \log_2 \left( 1 + \alpha_m \right) \\
		&- \alpha_m + \frac{ \left( 1 + \alpha_m \right) \gamma_m }{ 1 + \gamma_m} \big],
	\end{split}
\end{equation}
where \(\alpha_m \ge 0\) is an auxiliary variable corresponding to time slot \(m\). Note that $\alpha_m$ is introduced as an auxiliary variable to facilitate the convex reformulation of the optimization problem. Next, to decouple the product term involving \(\gamma_m\), we adopt the quadratic transformation technique, introducing another auxiliary variable \(\beta_m \in \mathbb{C}\). This yields
\begin{align}
	\scalebox{1.05}{$
	\frac{\left( 1 + \alpha_m \right) \gamma_m }{ 1 + \gamma_m} = 2 \sqrt{ \left(  1 + \alpha_m \right)  } \, \text{Re} \left(  \beta_m^\ast A_m \right) - \left| \beta_m \right|^2 B_m,
	$}
\end{align}
%
%
%
where \(A_m\) and \(B_m\) denote the signal and interference-related terms, respectively. They are defined as
\begin{equation}
	\begin{split}
		A_m &= \sqrt{P_\text{s} \bar{D}_{jk} \left( t_m \right) } \mathbf{s}_{jk} \left(\tilde{\mathbf{a}}_{jk} (t_m), \mathbf{c}\right)^\text{H} \mathbf{w} \left[ m \right], \\
		B_m &= P_\text{I} + \sigma^2 \left\|  \mathbf{w} \left[ m \right] \right\|_2^2,
	\end{split}
\end{equation}
where the term \(P_\text{I}\) captures the aggregate power of interference from other visible satellites, given by
\begin{align}
	P_\text{I} \triangleq P_\text{s} \sum\limits_{l \in \mathcal{K}_j \left( t_m \right) \atop j \in \mathcal{J} } \bar{D}_{jl} \left( t_m \right) \left| \mathbf{s}_{jl} \left(\tilde{\mathbf{a}}_{jl} (t_m), \mathbf{c}\right)^\text{H} \mathbf{w} \left[ m \right] \right|^2.
\end{align}
Substituting the transformed expressions into the objective, we obtain the following equivalent reformulation of the original problem (P1):
\begin{equation}
	\begin{split}
		(\text{P2}) \quad {\mathop{\max}\limits_{\mathbf{c}, \mathbf{w}, \bm{\alpha}, \bm{\beta}}} \quad & f \left( \mathbf{c}, \mathbf{w}, \bm{\alpha}, \bm{\beta} \right) \\
		s.t. \,\,\,\
		&\text{C1} \sim \text{C3}, \\
		&\text{C4}: \alpha_m \ge 0,
	\end{split}
\end{equation}
where \(\bm{\alpha} = \{\alpha_m\}_{m \in \mathcal{M}}\) and \(\bm{\beta} = \{\beta_m\}_{m \in \mathcal{M}}\) are the sets of auxiliary variables. The reformulated objective function is
\begin{equation}
	\begin{split}
		f \left( \mathbf{c}, \mathbf{w}, \bm{\alpha}, \bm{\beta} \right)& = \sum_{m \in \mathcal{M}} \left[ \log_2 \left( 1 + \alpha_m \right) - \alpha_m \right. \\
		&\left. + 2 \sqrt{ 1 + \alpha_m } \, \text{Re} \left( \beta_m^\ast A_m \right) - \left| \beta_m \right|^2 B_m \right].
	\end{split}
\end{equation} \par
Although the above transformation simplifies the original problem structure, the optimization problem (P2) remains non-convex due to the coupling between \(\mathbf{w}\), \(\mathbf{c}\), and the auxiliary variables. To tackle this challenge, we propose an iterative algorithm based on the BCD method. This method alternately updates each variable block—namely, \(\bm{\alpha}\), \(\bm{\beta}\), \(\mathbf{w}\), and \(\mathbf{c}\)—while keeping the others fixed. This decomposition breaks down the overall non-convex problem into several smaller subproblems that are easier to solve. Then, the solution is iteratively refined until convergence is reached. The following subsections detail the optimization procedures for each variable block.
\subsection{Optimization Design of AWV}
This subsection details the optimization of the auxiliary variables $\bm{\alpha}$ and $\bm{\beta}$, followed by updating the AWV $\mathbf{w}$ under fixed APV $\mathbf{c}$. The goal is to maximize the objective function in (P2) by sequentially solving two tractable subproblems. \par

\textit{1) Update of \(\bm{\alpha}\) and \(\bm{\beta}\):}  
Given fixed $\mathbf{w}$ and $\mathbf{c}$, the optimal auxiliary variables can be obtained in closed form by setting the partial derivatives of the objective function with respect to \(\alpha_m\) and \(\beta_m\) to zero, i.e., $\partial f / \partial \alpha_m = 0$ and $\partial f / \partial \beta_m = 0$. Solving these conditions yields
\begin{align} \label{eq1}
	\alpha_m^\star = \gamma_m \quad \text{and} \quad \beta_m^\star = \frac{\sqrt{ \left( 1 + \alpha_m \right) } A_m}{B_m}.
\end{align} \par
\textit{2) Update of \(\mathbf{w}\):}  
With $\bm{\alpha}$, $\bm{\beta}$, and $\mathbf{c}$ fixed, problem (P2) reduces to a quadratic minimization with respect to the AWV $\mathbf{w}$. Specifically, we arrive at the following problem:
%
%
%
\begin{equation}
	\begin{split}
		(\text{P3}) \ {\mathop{\min}\limits_{\mathbf{w}}} \ & \sum\limits_{m \in \mathcal{M}} \big[ \left| {\beta}_m \right|^2 B_m \\
		&\quad\quad\quad\quad - 2 \sqrt{ \left(  1 + \alpha_m \right)  } {\rm{Re}} \left\lbrace  {\beta}_m^\ast A_m \right\rbrace \big]\\
		s.t. \,\,\,\
		&\text{C5}: m \in \mathcal{M}.
	\end{split}
\end{equation}
For the objective of problem (P3), we can rewritten it as $\sum\limits_{m \in \mathcal{M}} \mathbf{w}^{\text{H}}[m] \mathbf{U}[m] \mathbf{w}[m] - 2 \text{Re} \left( \mathbf{w}^{\text{H}}[m] \mathbf{v}[m] \right)$,
%
%
where the Hermitian matrix \(\mathbf{U}[m]\) and vector \(\mathbf{v}[m]\) are respectively defined as
\begin{equation}
	\begin{split}
		&\mathbf{U} \left[ m \right]  = \left| \beta_m \right|^2 \big[ \sigma^2 \mathbf{I} + P_\text{s} \\
		& \times \sum\limits_{l \in \mathcal{K}_j \left( t_m \right) \atop j \in \mathcal{J}} \bar{D}_{jl} \left( t_m \right) \mathbf{s}_{jl} \left(\tilde{\mathbf{a}}_{jl} (t_m), \mathbf{c}\right) \mathbf{s}_{jl} \left(\tilde{\mathbf{a}}_{jl} (t_m), \mathbf{c}\right)^\text{H} \big],
	\end{split}
\end{equation}
and
\begin{align}
	\mathbf{v} \left[ m \right] = \sqrt{P_s \bar{D}_{jk} \left( t_m \right) \left( 1 + \alpha_m \right) } \beta_m \mathbf{s}_{jk} \left(\tilde{\mathbf{a}}_{jk} (t_m), \mathbf{c}\right).
\end{align}
This leads to the following convex quadratic optimization problem:
\begin{equation}
	\scalebox{0.92}{$
	\begin{split}
		(\text{P4}) \ {\mathop{\min}\limits_{\mathbf{w}}} \  &\sum\limits_{m \in \mathcal{M}} \mathbf{w}^{\text{H}} \left[ m \right] \mathbf{U} \left[ m \right] \mathbf{w} \left[ m \right] - 2 \text{Re} \left( \mathbf{w}^{\text{H}} \left[ m \right] \mathbf{v} \left[ m \right] \right) \\
		s.t. \,
		&\ \text{C5}.
	\end{split}
$}
\end{equation}
The closed-form optimal solution for each time slot \(m\) is obtained by setting the derivative of the objective to zero, resulting in
\begin{align} \label{eq2}
	\mathbf{w}^\star [m] = \mathbf{U}^{-1}[m] \mathbf{v}[m], \quad m \in \mathcal{M}.
\end{align}
This closed-form update greatly simplifies the AWV optimization and ensures computational efficiency in each iteration.
\subsection{Optimization Design of APV}
This subsection addresses the optimization of the ground station's APV \(\mathbf{c}\), while the variables \(\mathbf{w}\), \(\bm{\alpha}\), and \(\bm{\beta}\) are held fixed. The original objective in problem (P2) is first simplified by omitting constant terms that are irrelevant to \(\mathbf{c}\), resulting in the following reformulated subproblem:
\begin{equation}
	\begin{split}
		(\text{P5}) \quad {\mathop{\min}\limits_{\mathbf{c}}} \quad & \tilde{f} \left( \mathbf{c} \right) \\
		s.t. \quad & \text{C1} \sim \text{C3},
	\end{split}
\end{equation}
where the reduced objective \(\tilde{f}(\mathbf{c})\) is given by
\begin{align}
	\tilde{f}(\mathbf{c}) &= \sum_{m \in \mathcal{M}} \left| \beta_m \right|^2 B_m - 2 \sqrt{1 + \alpha_m} \, \text{Re}\left( \beta_m^\ast A_m \right).
\end{align}
Substituting the expressions for \(A_m\) and \(B_m\), we rewrite \(\tilde{f}(\mathbf{c})\) in a more explicit form:
\begin{equation} \label{eq31}
	\scalebox{0.87}{$
	\begin{split}
		&\tilde{f} \left( \mathbf{c}\right) = \sum_{m \in \mathcal{M}} \left| \beta_m \right|^2 P_\text{s} \\
		&\times \sum\limits_{l \in \mathcal{K}_j \left( t_m \right) \atop j \in \mathcal{J} } \big( \bar{D}_{jl} \left( t_m \right) \mathbf{s}_{jl} \left(\tilde{\mathbf{a}}_{jl} (t_m), \mathbf{c}\right)^\text{H} \mathbf{W} \left[ m \right] \mathbf{s}_{jl} \left(\tilde{\mathbf{a}}_{jl} (t_m), \mathbf{c}\right) \big) \\
		&\ \ - 2 \sqrt{ P_\text{s} \bar{D}_{jk} \left( t_m \right) \left( 1 + \alpha_m \right) } \text{Re} \big( \beta_m \mathbf{w} \left[ m \right]^\text{H} \mathbf{s}_{jk} \left(\tilde{\mathbf{a}}_{jk} (t_m), \mathbf{c}\right) \big),
	\end{split}
$}
\end{equation}
where \(\mathbf{W} \left[ m \right] \triangleq \mathbf{w} \left[ m \right] \mathbf{w} \left[ m \right]^\text{H}\), and the dependence of \(\mathbf{s}_{jl}\) and \(\mathbf{s}_{jk}\) on \(\mathbf{c}\) and \(t_m\) is omitted for brevity. \par
To deal with the non-convexity of $\tilde{f}(\mathbf{c})$, we adopt the SCA method and introduce two supporting lemmas inspired by \cite{10504625}. Specifically, Lemma~1 is directly adopted from \cite{10504625}, while Lemma~2 is a modified version that follows a similar approximation principle tailored to our model. \par
\textit{Lemma 1:}  
For a Hermitian matrix \(\mathbf{L}\), the quadratic form \(\mathbf{x}^\text{H} \mathbf{L} \mathbf{x}\) can be upper-bounded as
\begin{align} \label{eq32}
	\scalebox{0.91}{$
	\mathbf{x}^\text{H} \mathbf{L} \mathbf{x} \le \mathbf{x}^\text{H} \mathbf{M} \mathbf{x} + 2 \text{Re} \left( \mathbf{x}^\text{H} \left( \mathbf{L} - \mathbf{M} \right)  \mathbf{x}_0 \right) + \mathbf{x}_0^\text{H} \left( \mathbf{M} - \mathbf{L} \right) \mathbf{x}_0,
	$}
\end{align}
where \(\mathbf{M} \succeq \mathbf{L}\), and equality is attained at \(\mathbf{x} = \mathbf{x}_0\). \par
\textit{Lemma 2:} For the linear form $g \left( \mathbf{c} \right) = \text{Re} \left( \mathbf{b}^\text{H} \mathbf{s} \left( \mathbf{c} \right) \right) $, where $\mathbf{s} \left( \mathbf{c} \right) = \left[e^{j \mathbf{a}^\text{T} \mathbf{c}_n }\right]^\text{T}_{n \in \mathcal{N}}$ is a steering vector as the same definition in this paper, $\mathbf{c} = \left[\mathbf{c}_n \right]^\text{T}_{n \in \mathcal{N}}$, and $\mathbf{c}_n = \left[c_{n1}, c_{n2} \right]^\text{T}$, we can respectively obtain the upper bound and lower bound as
\begin{equation} \label{eq33}
	\begin{split}
		&g \left( \mathbf{c} \right) \le g \big( \mathbf{c}^{\left(i \right) } \big) + \nabla g \big( \mathbf{c}^{\left(i \right) } \big)^\text{T} \big( \mathbf{c} - \mathbf{c}^{\left(i \right) } \big) \\
		&\quad\quad\quad + \frac{1}{2} \left\| \mathbf{a}\right\|_2^2 \left\| \mathbf{b}\right\|_2 \big( \mathbf{c} - \mathbf{c}^{\left(i \right) } \big)^\text{T} \big( \mathbf{c} - \mathbf{c}^{\left(i \right) } \big),\\
		&g \left( \mathbf{c} \right) \ge g \big( \mathbf{c}^{\left(i \right) } \big) + \nabla g \big( \mathbf{c}^{\left(i \right) } \big)^\text{T} \left( \mathbf{c} - \mathbf{c}^{\left(i \right) } \right) \\
		&\quad\quad\quad - \frac{1}{2} \left\| \mathbf{a}\right\|_2^2 \left\| \mathbf{b}\right\|_2 \big( \mathbf{c} - \mathbf{c}^{\left(i \right) } \big)^\text{T} \big( \mathbf{c} - \mathbf{c}^{\left(i \right) } \big).
	\end{split}
\end{equation}
where \(\mathbf{c}^{\left(i \right)}\) represents the value of \(\mathbf{c}\) at the \(i\)th iteration. Assuming that \(\mathbf{a} = [a_1, a_2]^\text{T}\) and \(\mathbf{b} = \left[ \left| b_n \right| e^{j\theta_n} \right]^\text{T}_{n \in \mathcal{N}}\), we have $\nabla g \big( \mathbf{c}^{\left(i \right) } \big) = \left[ -a_{\lceil n / 2 \rceil}
\left| b_n\right| \sin \left( \mathbf{a} \mathbf{c}_n - \theta_n \right) \right]^\text{T}_{n \in \bar{\mathcal{N}}}$,
%
%
where $\lceil \cdot \rceil$ represents the ceiling function, which rounds a number up to the nearest integer, and $\bar{\mathcal{N}} \triangleq \left\lbrace 1, \cdots, 2N \right\rbrace $. \par
Applying Lemma~1 and Lemma~2 to the objective function $\tilde{f}(\mathbf{c})$, we then construct a convex surrogate function at the $i$th iteration for the SCA method. Specifically, at the current iterate $\mathbf{c}^{(i)}$, the objective function in (P5) is upper-bounded as
\begin{equation} \label{eq34}
	\begin{split}
		&\tilde{f} \left( \mathbf{c}\right)\le \sum_{m \in \mathcal{M}} \text{Re} \big( \sum\limits_{l \in \mathcal{K}_j \left( t_m \right) \atop j \in \mathcal{J} } \mathbf{z}_l \left[ m \right]^\text{H} \mathbf{s}_{jl} \left(\tilde{\mathbf{a}}_{jl} (t_m), \mathbf{c}\right) \\
		& \quad\quad\quad\quad- \mathbf{z} \left[ m \right]^\text{H} \mathbf{s}_{jk} \left(\tilde{\mathbf{a}}_{jk} (t_m), \mathbf{c}\right) \big) + \text{constant}\ 1,\\
		&\le \sum_{m \in \mathcal{M}} \sum\limits_{l \in \mathcal{K}_j \left( t_m \right) \atop j \in \mathcal{J} } \frac{1}{2} \left\| \tilde{\mathbf{a}}_{jl} (t_m) \right\|_2^2 \left\| \mathbf{z}_l \left[ m \right] \right\|_2 \left\| \mathbf{c} \right\|^2 \\
		&+ \big( \nabla \text{Re} \big( \mathbf{z}_l \left[ m \right]^\text{H} \mathbf{s}_{jl} \left(\tilde{\mathbf{a}}_{jl} (t_m), \mathbf{c}^{\left( i\right) }\right) \big) \\
		&\quad\quad\quad\quad\quad\quad\quad\quad - \left\| \tilde{\mathbf{a}}_{jl} (t_m) \right\|_2^2 \left\| \mathbf{z}_l \left[ m \right] \right\|_2 \mathbf{c}^{\left( i\right) } \big)^\text{T} \mathbf{c} \\
		&+ \frac{1}{2} \left\| \tilde{\mathbf{a}}_{jk} (t_m) \right\|_2^2 \left\| \mathbf{z} \left[ m \right] \right\|_2 \left\| \mathbf{c} \right\|^2 \\
		&\quad\quad\quad\quad\quad\quad - \big( \nabla \text{Re} \big( \mathbf{z} \left( m \right)^\text{H} \mathbf{s}_{jk} \left(\tilde{\mathbf{a}}_{jk} (t_m), \mathbf{c}^{\left( i\right) }\right) \big) \\
		& + \left\| \tilde{\mathbf{a}}_{jk} (t_m) \right\|_2^2 \left\| \mathbf{z} \left[ m \right] \right\|_2 \mathbf{c}^{\left( i\right) } \big)^\text{T} \mathbf{c} + \text{constant}\ 2 \triangleq \bar{f} \left( \mathbf{c}\right),
	\end{split}
\end{equation}
where
%
%
%
\begin{equation} \label{eq35} \scalebox{0.95}{$
	\begin{split}
		\mathbf{z}_l \left[ m \right] &= 2 \left| \beta_m \right|^2 \bar{D}_{jl} \left( t_m \right) \\
		&\quad P_\text{s} \big( \mathbf{W} \left[ m \right] - \left\| \mathbf{w} \left[ m \right] \right\|^2 \mathbf{I}_N \big) \mathbf{s}_{jl} \big(\tilde{\mathbf{a}}_{jl} (t_m), \mathbf{c}^{\left( i \right) } \big), \\
		\mathbf{z} \left[ m \right] & = 2 \beta_m^\ast \sqrt{\left( 1+\alpha_m\right) \bar{D}_{jk} \left( t_m \right) P_\text{s}} \mathbf{w} \left[ m \right],
	\end{split} $}
\end{equation} \par
Additionally, to handle the non-convex constraint C2, we linearize it using the first-order Taylor expansion at \(\mathbf{c}^{(i)}\) as
%
%
%
\begin{equation} \label{eq36} \scalebox{0.95}{$
	\begin{split}
		&\left\| \mathbf{c}_n - \mathbf{c}_{\tilde{n}} \right\|_2  \ge \big\| \mathbf{c}_n^{\left( i \right) } - \mathbf{c}_{\tilde{n}}^{\left( i \right) } \big\|_2 \\
		&\quad\quad\quad\ + \frac{\big( \mathbf{c}_n^{\left( i \right) } - \mathbf{c}_{\tilde{n}}^{\left( i \right) }\big)^{\text{T}} }{\big\| \mathbf{c}_n^{\left( i \right) } - \mathbf{c}_{\tilde{n}}^{\left( i \right) } \big\|_2} \big[ \left( \mathbf{c}_n - \mathbf{c}_{\hat{n}} \right) - \big( \mathbf{c}_n^{\left( i \right) } - \mathbf{c}_{\hat{n}}^{\left( i \right) } \big) \big]\\
		&\quad\quad\quad\ = \frac{\left( \mathbf{c}_n^{\left( i \right) } - \mathbf{c}_{\tilde{n}}^{\left( i \right) }\right)^{\text{T}}\left( \mathbf{c}_n - \mathbf{c}_{\hat{n}} \right)}{\big\| \mathbf{c}_n^{\left( i \right) } - \mathbf{c}_{\tilde{n}}^{\left( i \right) } \big\|_2}. 
	\end{split} $}
\end{equation}
Based on the above deduction, tackling problem (P5) can be transformed to solving the follwoing problem:
\begin{equation}
	\begin{split}
		(\text{P6}) \quad {\mathop{\min}\limits_{\mathbf{c}}} \quad & \bar{f} \left( \mathbf{c}\right) \\
		s.t. \,\,\,\
		&\text{C1}, \ \text{C3}, \\
		&\text{C6}: \frac{\big( \mathbf{c}_n^{\left( i \right) } - \mathbf{c}_{\tilde{n}}^{\left( i \right) }\big)^{\text{T}}\left( \mathbf{c}_n - \mathbf{c}_{\hat{n}} \right)}{\big\| \mathbf{c}_n^{\left( i \right) } - \mathbf{c}_{\tilde{n}}^{\left( i \right) } \big\|_2} \ge d_{\text{min}},
	\end{split}
\end{equation}
where constraint C6 is the linearized version of C2 as described above. By solving problem (P6) iteratively at each SCA iteration, the APV \(\mathbf{c}\) is gradually refined, ensuring convergence to a stationary point of the original problem. To summarize, the objective function $\tilde{f}(\mathbf{c})$ of problem (P5), given in \eqref{eq31}, is first upper-bounded by applying Lemma 1, as shown in \eqref{eq32}, where the equality can be attained. Then, following the derivations in Appendix A and the closed-form expression in Appendix B of \cite{10243545}, the upper and lower bounds in \eqref{eq33} are obtained based on Lemma 2. These inequalities satisfy the Taylor expansion theorem, and accordingly, the upper bound of $\tilde{f}(\mathbf{c})$ is derived as expressed in \eqref{eq34}. Furthermore, the constraint C2 in problem (P5) is transformed into C6 via the first-order Taylor expansion in \eqref{eq36}. Based on the Taylor theorem, problem (P5) can be relaxed into problem (P6), and the optimal solution of problem (P6) converges to a suboptimal solution of problem (P5).
\subsection{Convergence and Computational Complexity Analysis}
\begin{algorithm}[t]
	\caption{Overall algorithm for problem (P1)}  
	\label{algorithm1}
	\textbf{Input:} $K$, $J$, $R$, $H$, $\beta$, $T$, $T_\text{E}$, $\Theta_\text{u}$, $\lambda$, $\eta$, $r$, $M$, $\mathcal{A}$, $d_\text{min}$, $\epsilon$, $I_\text{max}$ \\
	\textbf{Output:} $\mathbf{w}^\star$, $\mathbf{c}^\star$.
	\begin{algorithmic}[1]
		\State Initialize $\mathbf{c}^{(0)}$ and $\mathbf{w}^{(0)}$ through Eq.\eqref{eq3}
		\For{$i = 1 : 1 : I_{\max}$}
		\State Update $\alpha_m^\star$ and $\beta_m^\star$ through Eq.\eqref{eq1}
		\For{$m = 1 : 1 : M$}
		\State Update $\mathbf{w}^{(i)}[m]$ according to Eq.\eqref{eq2}
		\EndFor
		\State Update $\mathbf{c}^{(i)}$ by solving problem (P6)
		\If{$| \bar{C} \left( \mathbf{c}^{(i)}, \mathbf{w}^{(i)} \right) - \bar{C} \left( \mathbf{c}^{(i-1)}, \mathbf{w}^{(i-1)} \right) | \leq \epsilon$}
		\State \textbf{break}
		\EndIf
		\EndFor
		\State Set AWV and APV as $\mathbf{w}^\star = \mathbf{w}^{(i)}$ and $\mathbf{c}^\star = \mathbf{c}^{(i)}$
		\State \Return $\mathbf{w}^\star$, $\mathbf{c}^\star$
	\end{algorithmic}
\end{algorithm}
Algorithm~\ref{algorithm1} presents the overall procedure for solving problem (P1), where the AWV $\mathbf{w}$ and the APV $\mathbf{c}$ are alternately optimized in an iterative manner. The AWV is initialized as the normalized channel vector between the ground station and the selected serving satellite $\text{S}_{jk}$, given by
\begin{align} \label{eq3}
	\mathbf{w}^{(0)}[m] = \frac{\mathbf{h}_{jk} \left( \mathbf{c}^{(0)}, t_m \right) }{ \left\| \mathbf{h}_{jk} \left( \mathbf{c}^{(0)}, t_m \right) \right\|_2 }, \quad m \in \mathcal{M}.
\end{align}
The initial APV, denoted as $\mathbf{c}^{(0)}$, is configured by uniformly distributing the antenna elements within the feasible region $\mathcal{A}$. In each iteration, the auxiliary variables, APV, and AWV are updated sequentially in lines 3, 4–5, and 6 of the algorithm, respectively. The algorithm terminates when the improvement of the average achievable rate $\bar{C}(\mathbf{c}, \mathbf{w})$ falls below a predefined threshold $\epsilon$, or the maximum number of iterations $I_{\max}$ is reached. In addition, the closed-form expressions in \eqref{eq1} and \eqref{eq2} are derived based on first-order derivatives, while the surrogate objective function $\bar{f}(\mathbf{c})$ in problem (P6) is constructed using Taylor series expansion. Since each update step solves a convex subproblem and the objective value is non-decreasing and upper-bounded, the value sequence generated by the algorithm is guaranteed to converge to a stationary objective value. Therefore, the convergence of Algorithm~\ref{algorithm1} is ensured. \par
Furthermore, the computational complexity of Algorithm~\ref{algorithm1} is primarily dominated by computing the solution to problem (P6). Problem (P6) is a convex quadratic program with $2N$ real-valued optimization variables, corresponding to the two-dimensional positions of $N$ antenna elements. The objective function $\bar{f}(\mathbf{c})$ is convex and quadratic, while the constraints consist of a bounded movement region for each antenna in C1 and a set of linearized minimum distance constraints between antenna pairs in C6. Since the number of inter-antenna distance constraints scales as $\mathcal{O}(N^2)$, problem (P6) involves $\mathcal{O}(N)$ variables and $\mathcal{O}(N^2)$ constraints. When solved using an interior-point method, the per-iteration computational complexity is on the order from $\mathcal{O}(N^3)$ to $\mathcal{O}(N^{3.5})$, depending on the number of active constraints. Therefore, in the worst-case scenario, the overall complexity of Algorithm~\ref{algorithm1} is $\mathcal{O}(I_\text{max} N^{3.5})$.
\section{Simulation Results}
This section analyzes the achievable rate performance of the proposed MA-assisted LEO satellite system under various antenna configuration schemes. In particular, we consider two baseline schemes based on FPA arrays to benchmark the performance of our MA scheme. These comparisons aim to highlight the advantages of enabling antenna spatial configuration at the ground station. The considered schemes are summarized as follows:
\begin{itemize}
	\item MA: This is the proposed movable antenna scheme, which jointly optimizes the AWVs and APV at the ground station to solve (P1). Each antenna element can move within a predefined two-dimensional region \(\mathcal{A}\). The joint optimization of AWVs and APV enables the ground station to dynamically adapt its array geometry, aligning with the service satellite signal directions and suppressing interference signals from other satellites.
	
	\item SFPA: The sparse FPA (SFPA) scheme serves as a baseline that assumes the antenna elements at the ground station are statically and uniformly distributed within the region \(\mathcal{A}\), with relatively large inter-element spacing. The AWVs are optimized based on the fixed geometry. This scheme serves as a reference for sparse array performance.
	
	\item DFPA: The dense FPA (DFPA) scheme also assumes static antenna placement, but with the antenna elements tightly packed using the minimum allowable spacing (i.e., \(0.5\lambda\)) within the movement region \(\mathcal{A}\). Similar to the SFPA scheme, only the AWVs at the ground station are optimized for the given array layout. This scheme serves as a reference for dense array performance.
\end{itemize}
\begin{table}[t]
	\footnotesize
	\begin{center}
		\caption{Simulation Parameters and Values}\label{T1}~~\\
		\begin{tabular}{c|c|c}
			\hline
			Variable & Definition & Value\\ \hline \hline
			$H$ & Altitude of satellite & $550$ km\\ \hline
			$R$ & Earth radius & $6371$ km \\ \hline
			$\beta$ & Orbital inclination angle & $65^\circ$ \\ \hline
			$T_{\text{E}}$ & Earth rotation period & $15 T$\\ \hline
			$M$ & Number of time slots & $500$ \\ \hline
			$f_\text{c}$ & Carrier frequency & $14$ GHz\\ \hline
			$d_\text{min}$ & Inter-antenna distance & $0.5\lambda$ \\ \hline
			$\mathcal{A}$ & Antenna moving region & $3\lambda \times 3\lambda$ \\ \hline
			$\sigma^2$ & Noise power & $-120$ dBm \\ \hline
			$N$ & Number of antennas & $16$ \\ \hline
			$P_\text{s}$ & Satellite transmit power & $30$ dBW \\ \hline
			$r$ & Satellite antenna’s circular aperture radius & $3 \lambda$ \\ \hline
			$\eta$ & Aperture efficiency & $0.5$ \\ \hline
		\end{tabular}
	\end{center}
\end{table}
The simulation parameters used for performance evaluation are summarized in Table~\ref{T1}. These parameters cover a comprehensive set of system aspects, including satellite deployment, signal characteristics, and antenna design constraints, and are chosen to reflect practical LEO satellite communication scenarios in accordance with existing standards and literature. \par
\begin{figure}
	\includegraphics[width=0.45\textwidth]{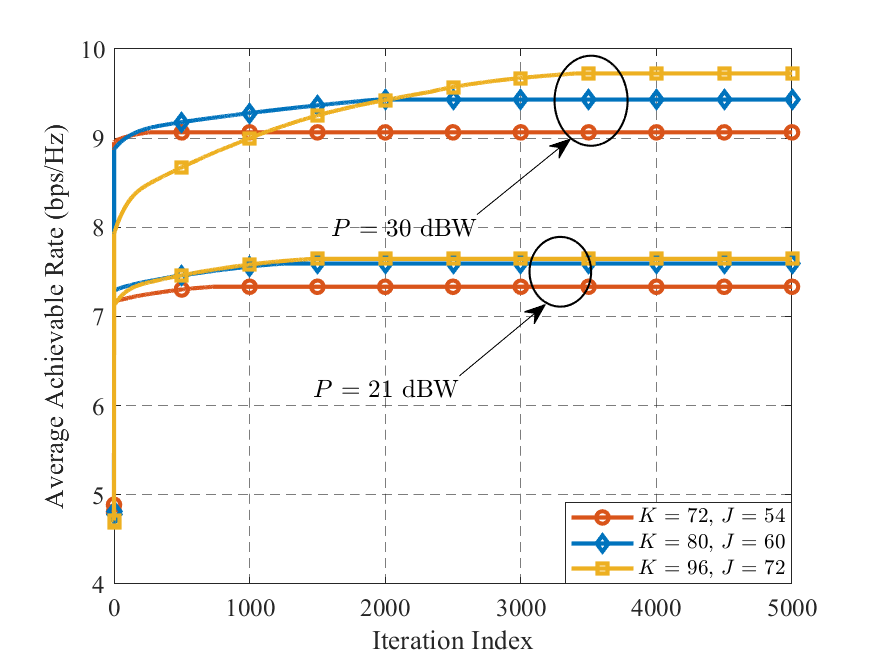}
	\caption{The rate convergence performance of the proposed algorithm for different parameters.}
	\label{figure1}
\end{figure}
\begin{figure}
	\includegraphics[width=0.45\textwidth]{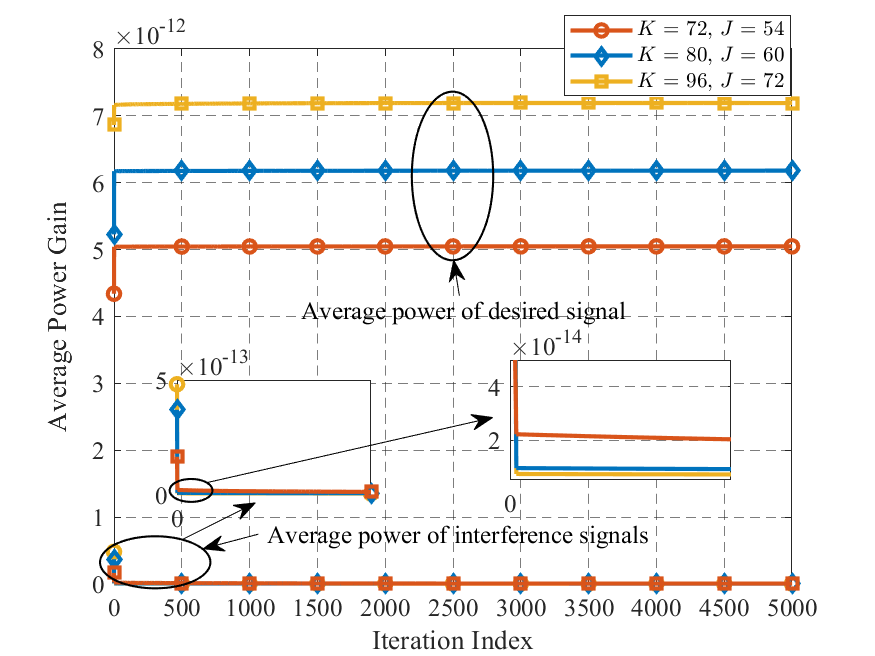}
	\caption{The power gain convergence performance of the proposed algorithm for different parameters.}
	\label{figure1.2}
\end{figure}

Fig.~\ref{figure1} illustrates the convergence performance of the proposed optimization algorithm for the MA-assisted LEO satellite internet network. The number of satellites per orbit is set to \( K = 72 \), \( 80 \), and \( 96 \), while the number of orbits is configured as \( J = 54 \), \( 60 \), and \( 72 \), respectively. Following \cite{10436074} and \cite{9351765}, the numbers of satellites and orbits are set to reflect practical LEO systems and verify the proposed scheme under such conditions. The transmit power at each satellite is set to \( P = 21 \) and \( 30 \) dBW, and the iteration accuracy is set to \( \epsilon = 10^{-4} \). The results demonstrate that the proposed MA scheme consistently improves the average sum rate across all parameter settings. Additionally, increasing the satellite transmit power provides a rate gain, which is further enhanced by optimizing the antenna position. Furthermore, it is observed that the average achievable rate decreases as the number of satellites and orbits increases. This is because a larger number of satellites and orbits result in higher interference. However, the rate improvement achieved by the MA scheme becomes more pronounced as both the numbers of satellites and orbits increase, highlighting its effectiveness in mitigating interference. Moreover, after applying the MA scheme, the achievable rate in scenarios with more satellites and orbits surpasses that in scenarios with fewer satellites and orbits. This is because, in addition to suppressing interference, the MA scheme enhances the received signal quality, enabling the ground station to connect to a better service satellite during the considered period. This conclusion is further supported by Fig.~\ref{figure1.2}, which illustrates the average power gains of the desired signal and interference signals. It can be observed that the average gain of the desired signal increases with the number of satellites and orbits. This is because a denser constellation provides more satellite options, allowing the ground station to connect to a service satellite with a stronger link gain. By employing the MA scheme, the gain of the desired signal is further enhanced, while the interference signals are effectively suppressed. This dual improvement contributes to the observed increase in the average achievable rate under dense satellite deployments. \par
\begin{figure*}[t]
	\centering
	\subfigure[\texttt{MA}, $m = 40$]{%
		\includegraphics[width=0.31\textwidth]{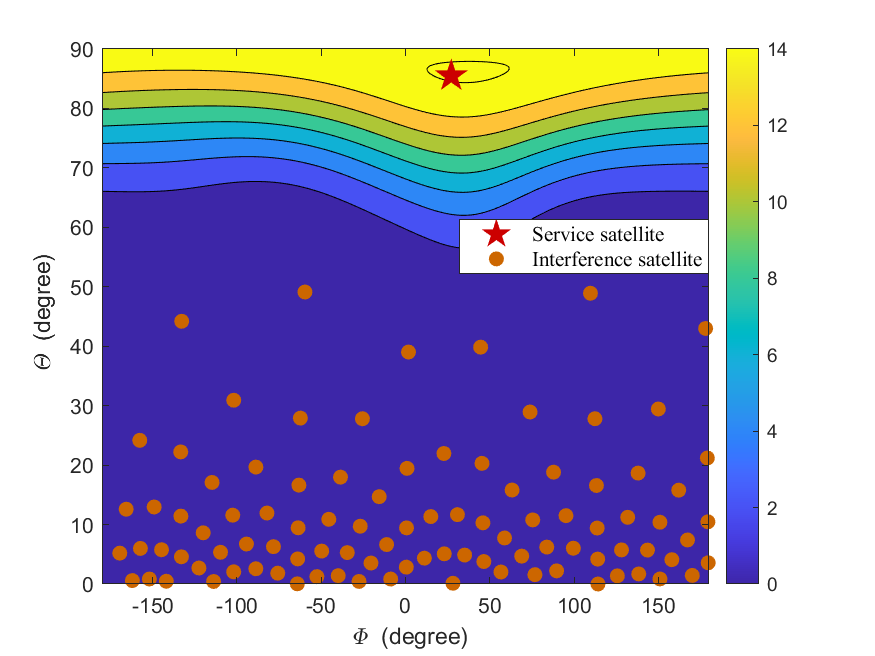}}
	\hfill
	\subfigure[\texttt{DFPA}, $m = 40$]{%
		\includegraphics[width=0.31\textwidth]{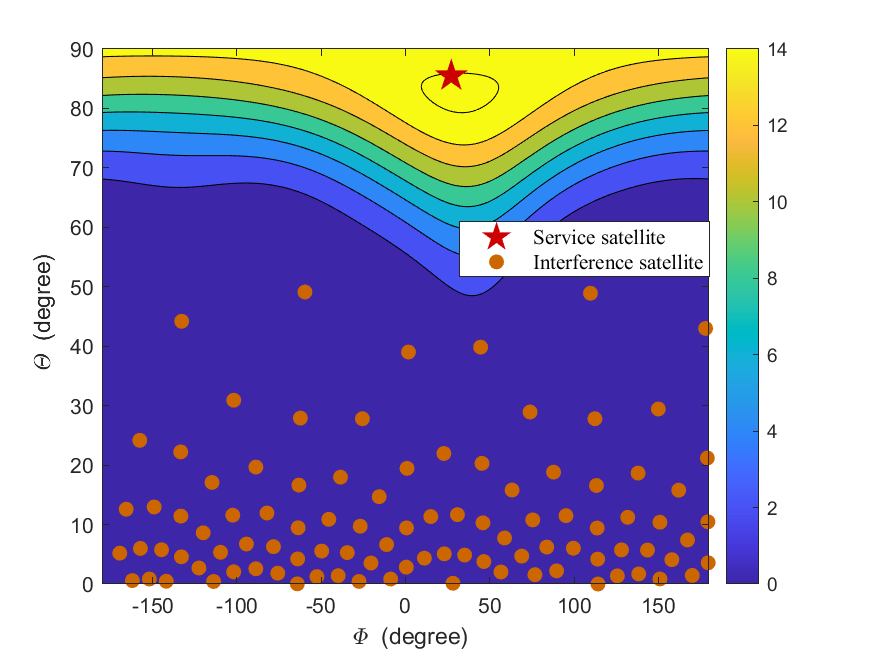}}
	\hfill
	\subfigure[\texttt{SFPA}, $m = 40$]{%
		\includegraphics[width=0.31\textwidth]{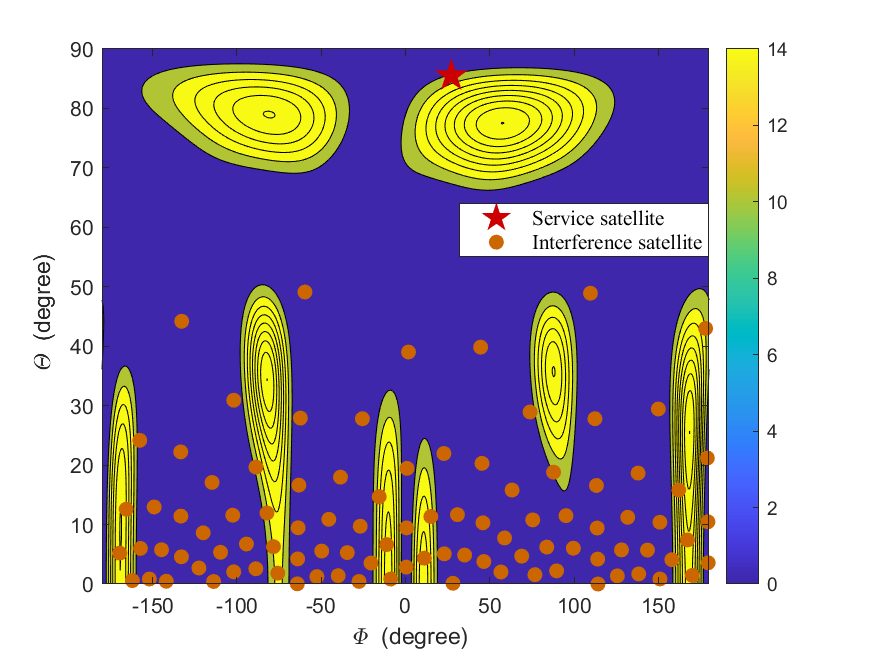}}
	
	\subfigure[\texttt{MA}, $m = 260$]{%
		\includegraphics[width=0.31\textwidth]{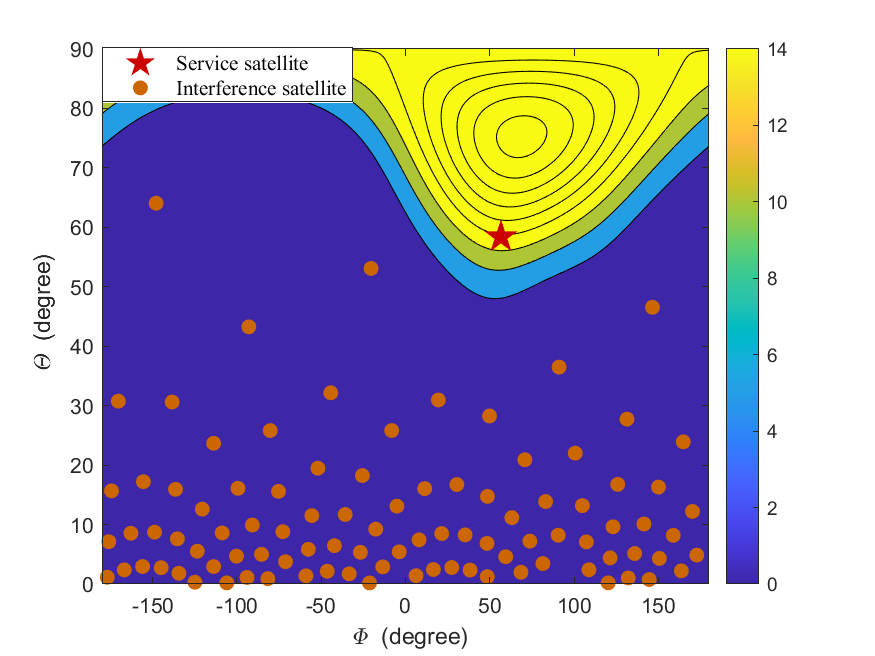}}
	\hfill
	\subfigure[\texttt{DFPA}, $m = 260$]{%
		\includegraphics[width=0.31\textwidth]{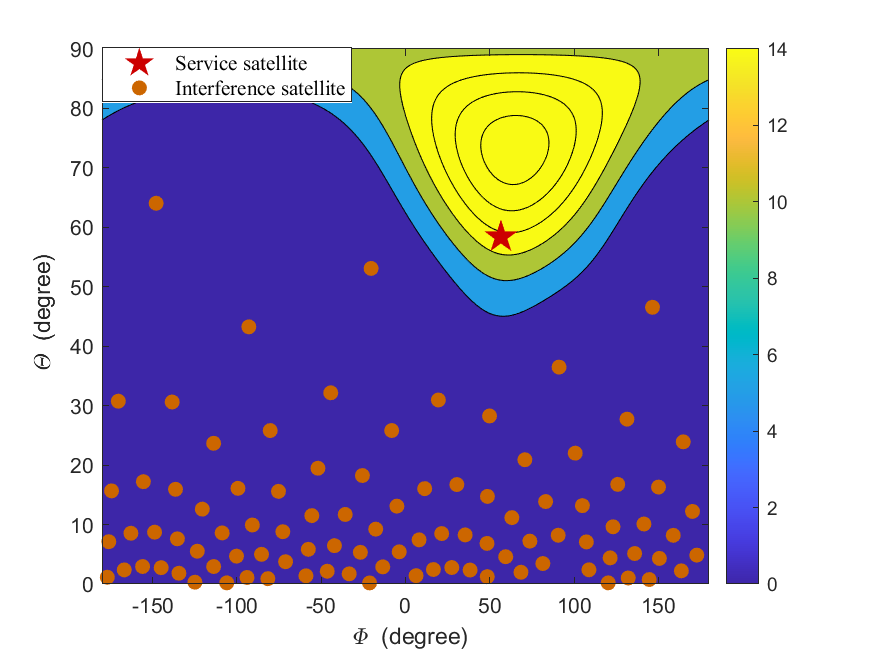}}
	\hfill
	\subfigure[\texttt{SFPA}, $m = 260$]{%
		\includegraphics[width=0.31\textwidth]{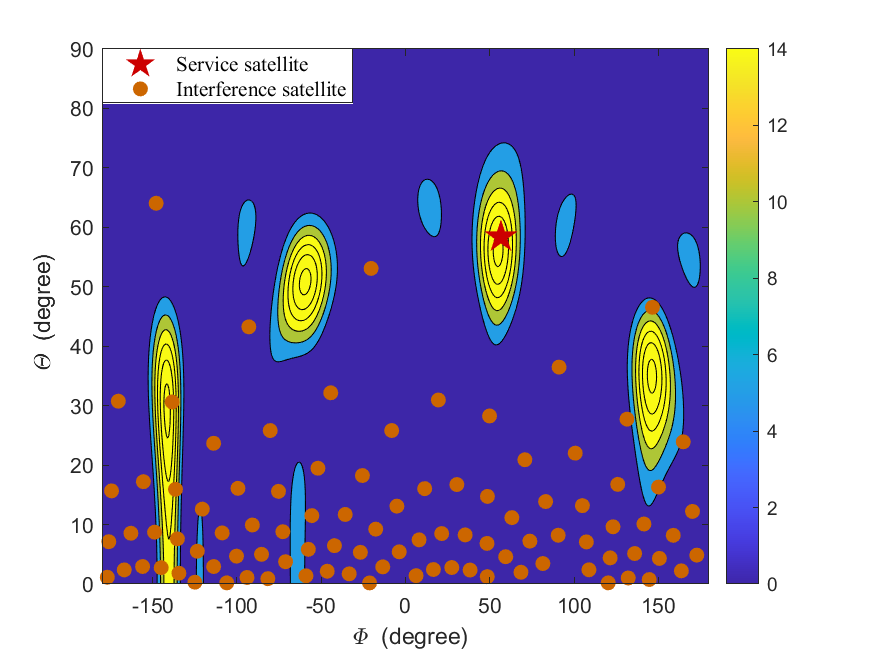}}
	
	\subfigure[\texttt{MA}, $m = 470$]{%
		\includegraphics[width=0.31\textwidth]{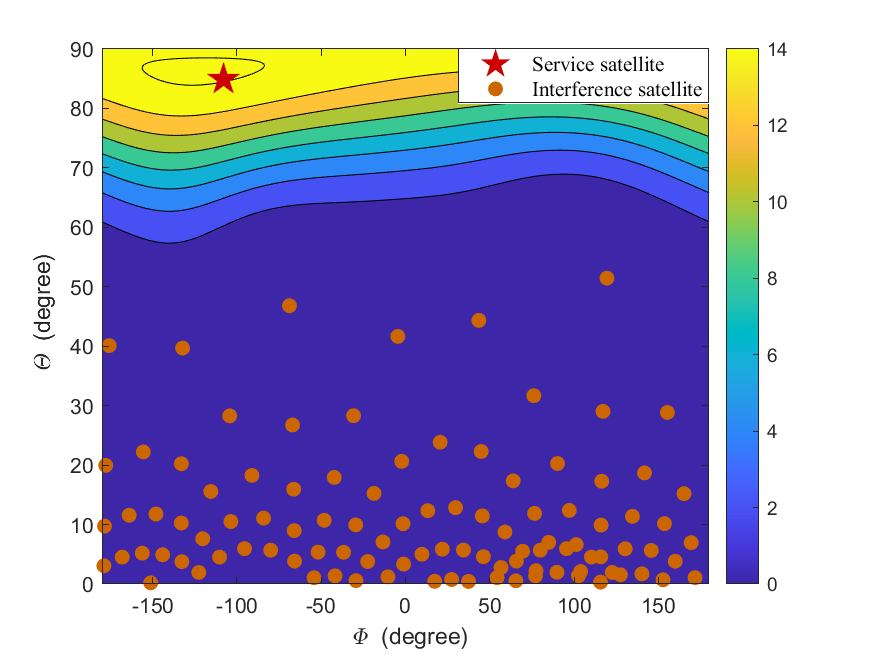}}
	\hfill
	\subfigure[\texttt{DFPA}, $m = 470$]{%
		\includegraphics[width=0.31\textwidth]{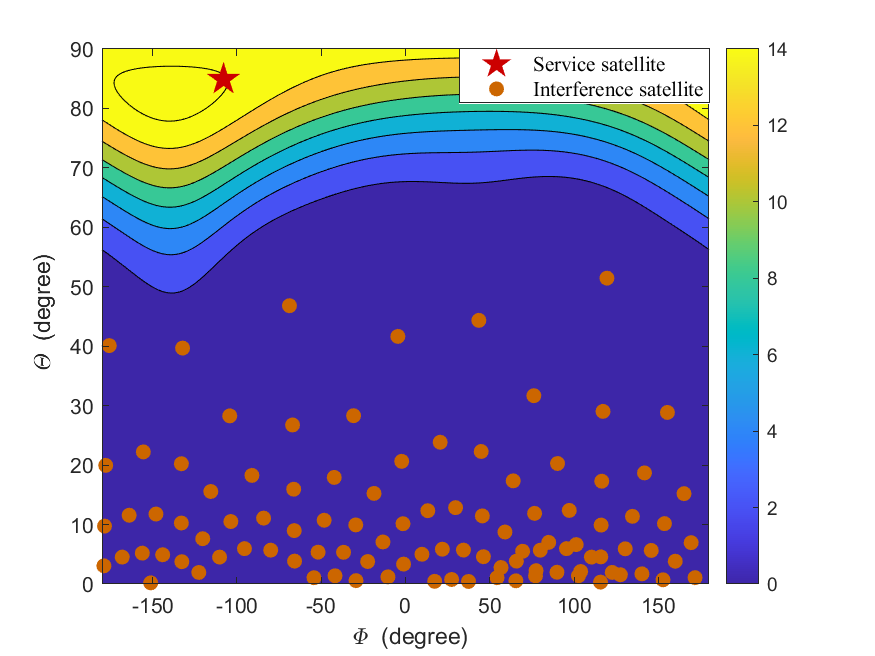}}
	\hfill
	\subfigure[\texttt{SFPA}, $m = 470$]{%
		\includegraphics[width=0.31\textwidth]{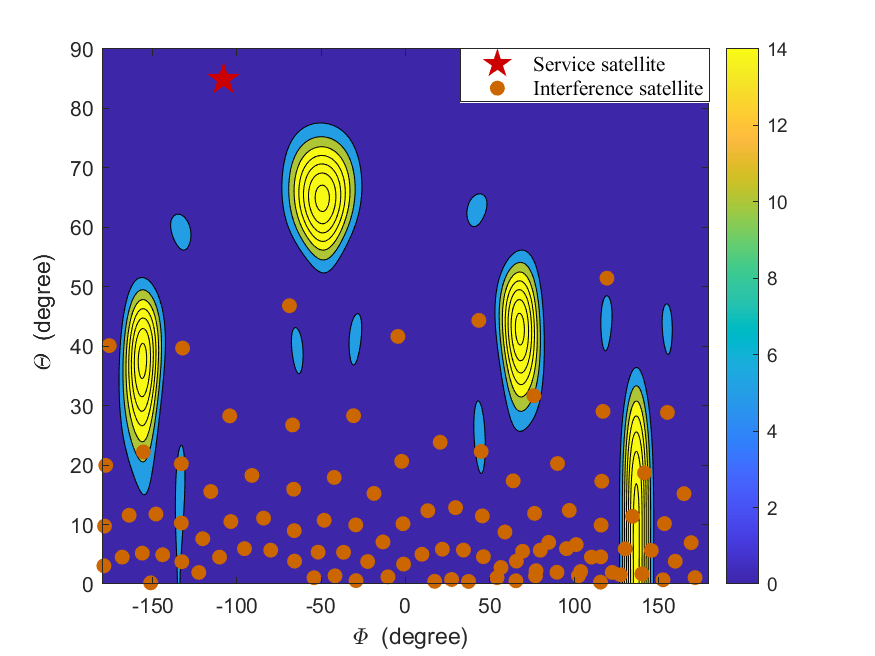}}
	
	\caption{The beamforming gain patterns for different optimization schemes and time slots.}
	\label{fig:beampatterns}
\end{figure*}
\begin{figure*}[t]
	\centering
	\includegraphics[width=0.93\textwidth]{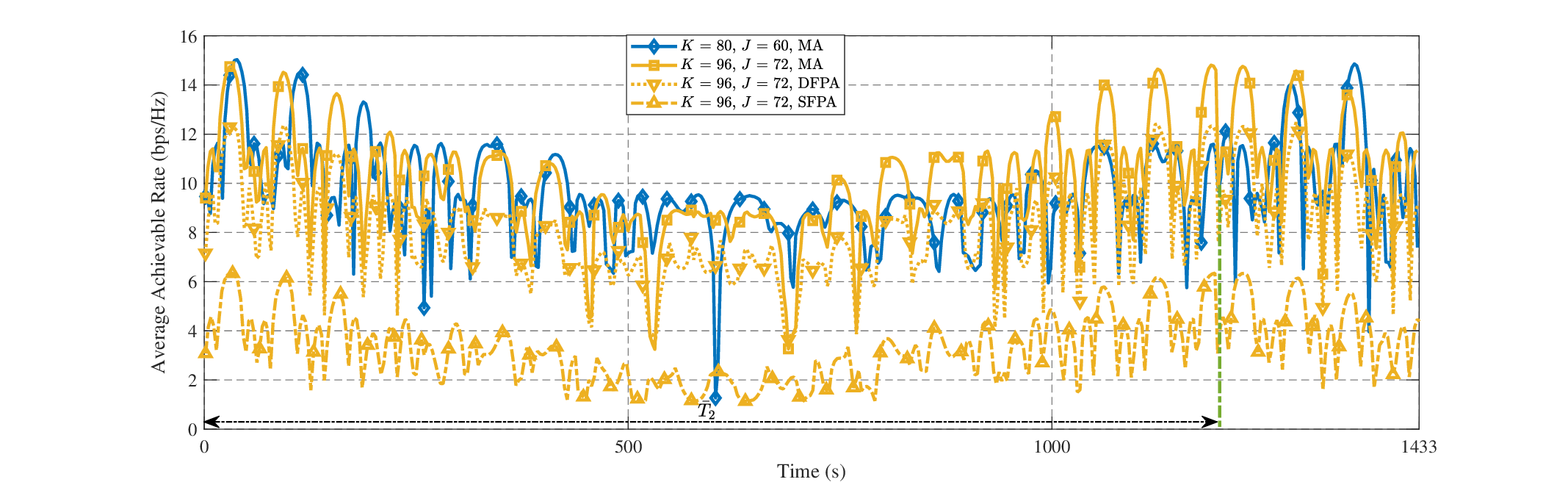}
	\caption{The achievable rates with different schemes at each time slot for different parameters.}
	\label{figure2}
\end{figure*}

Fig.~\ref{fig:beampatterns} shows the beamforming gain patterns for different optimization schemes and time slots, where the number of satellites per orbit is set to \( K = 96 \), and the number of orbits is configured as \( J = 72 \). The transmit power at each satellite is set to \( P = 30 \) dBW, and the time slots are set to \( m = 40 \), \( m = 260 \), and \( m = 470 \). For the SFPA optimization scheme, it can be observed that the beamforming gains of the interference satellites are higher than those in the other two schemes, thereby yielding the lowest achievable rate. In the DFPA optimization scheme, the service satellite is positioned closer to the main lobe compared with the SFPA scheme, while the interference satellites are located further away from the main lobe. This indicates a higher beamforming gain for the desired signal, thus improving the achievable rate. Furthermore, in the proposed MA scheme, which jointly designs the antenna positions and beamformer, the service satellite is positioned even closer to the main lobe compared with the DFPA scheme. In addition, the beam in the MA scheme is narrower than that in the DFPA scheme, allowing the ground station to achieve more effective beamforming for both signal reception and interference suppression. Moreover, it is observed that the achievable rate is the lowest at the time slot \( m = 260 \) compared with the other two time slots, even under the MA schemes. This indicates that the rate gain achieved by the MA optimization scheme is reduced at \( m = 260 \). The reason is that the geometry of the connected service satellite and visible interference satellites is less favorable at this time slot, while the proposed MA scheme aims to maximize the average achievable rate across all time slots rather than optimizing the rate at each individual time slot. The detailed achievable rates for each time slot are shown in Fig.~\ref{figure2}. \par
Fig.~\ref{figure2} depicts the achievable rates of the MA and two FPA schemes at different time instances, where the orbital period is set to \( T_E = 85952 \)~s. The number of satellites per orbit is set to \( K = 80 \) and \( 96 \), while the number of orbits is configured as \( J = 60 \) and \( 72 \), respectively. The optimization periods, calculated as \( T_E / J \), are given as \( \bar{T}_1 = 1433 \) and \( \bar{T}_2 = 1194 \), respectively. The value of \( \bar{T}_1 \) corresponds to the range of the x-axis, while \( \bar{T}_2 \) is indicated in Fig.~\ref{figure2}. For the SFPA scheme, the achievable rate is consistently lower than that of the DFPA scheme under the same parameter settings. Moreover, the MA scheme further improves the achievable rate by optimizing the antenna position. Focusing on the MA scheme with \( K = 96 \) and \( J = 72 \), it is observed that during the initial time period, the service satellite is positioned nearer to the ground station, resulting in a higher achievable rate compared to later time instances. Nevertheless, as the next satellite approaches and becomes the new serving satellite, the achievable rate increases again, showing a recurring pattern approximately every \( 1194 \)~s. Additionally, while increasing the number of satellites and orbits introduces more interference, it also expands the set of available satellites, enabling the ground station to select a serving satellite with higher path gains. This explains why, in some time slots, the achievable rate with a denser satellite constellation (more satellites and orbits) surpasses that of a less dense satellite constellation. Furthermore, even considering the Earth's rotation, the achievable rate variations exhibit an approximately periodic pattern, demonstrating that the MA scheme maintains its performance over successive cycles. \par
\begin{figure}
	\includegraphics[width=0.45\textwidth]{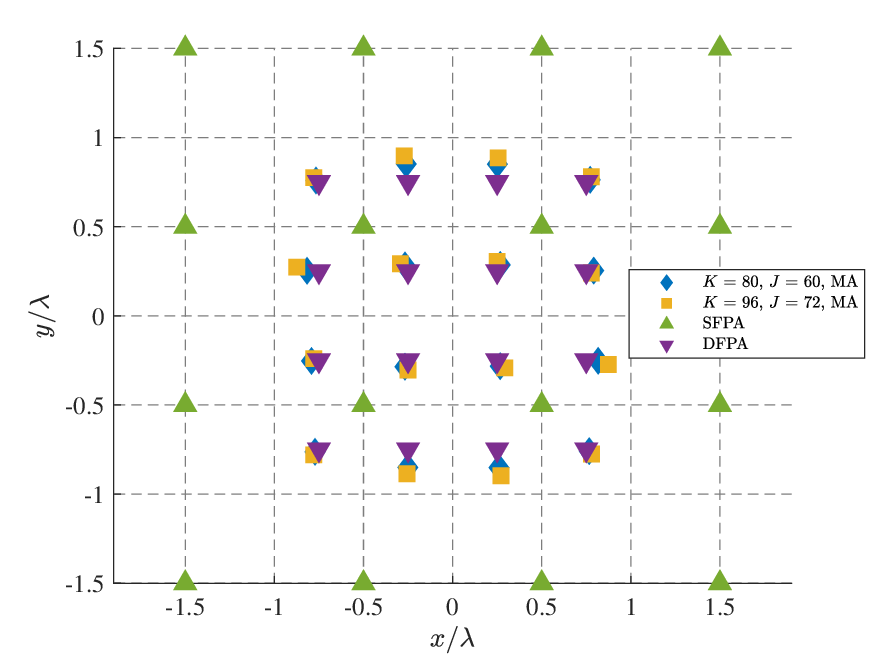}
	\caption{\footnotesize{The optimzed antenna positions with the proposed MA scheme for different parameters.}}
	\label{figure3}
\end{figure}
Fig.~\ref{figure3} illustrates the optimized and fixed antenna positions at the ground station for different parameter settings, where \( K = 80 \) and \( 96 \), and \( J = 60 \) and \( 72 \), respectively. Under the SFPA scheme, the ground station's antennas are evenly spaced at an interval of \( \lambda \) within a range of \( 3\lambda \times 3\lambda \), while under the DFPA scheme, the antennas are placed with an interval of \( 0.5\lambda \) within a range of \( 1.5\lambda \times 1.5 \lambda \). In contrast, the MA scheme yields different optimized antenna positions depending on the number of satellites and orbits. Furthermore, as the satellite density increases (i.e., more satellites and orbits), the optimized antenna positions shift outwards to enhance angular resolution, thereby improving both interference suppression and signal reception. These optimized positions enable the ground station to achieve varying levels of interference mitigation while connecting to the optimal service satellite. Thus, the proposed MA scheme can adaptively configure antenna positions based on the satellite network conditions, offering a more effective and flexible antenna initialization strategy. \par
\begin{figure}
	\includegraphics[width=0.45\textwidth]{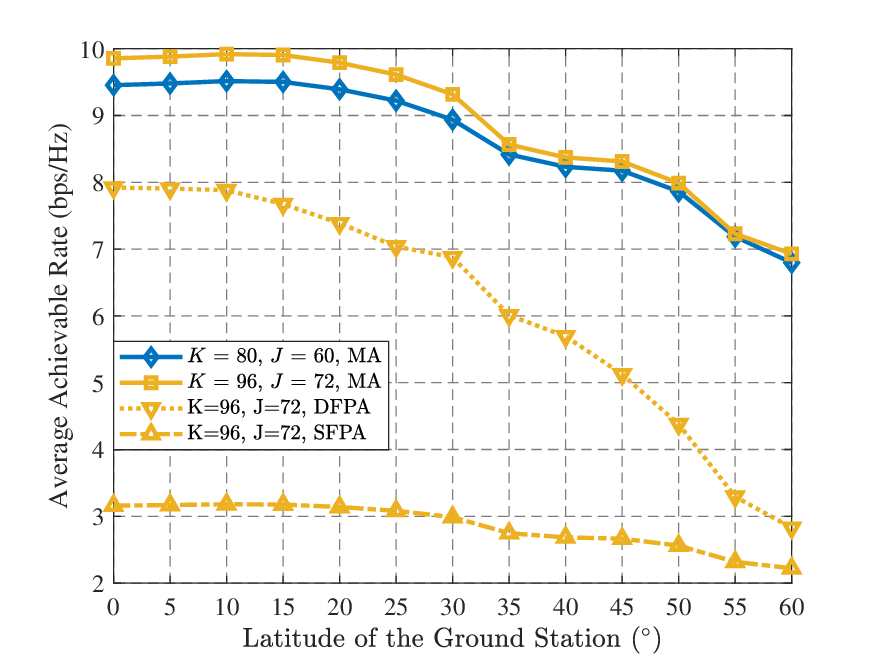}
	\caption{\footnotesize{The ground station's latitude versus its average achievable rate for different schemes and parameters.}}
	\label{figure4}
\end{figure}
Fig.~\ref{figure4} compares the average achievable rate under different latitudes of the ground station, where the number of satellites per orbit is set to \( K = 80 \) and \( 96 \), and the number of orbits is configured as \( J = 60 \) and \( 72 \), respectively. The results indicate that the average achievable rate fluctuates as the ground station's position changes, primarily due to variations in the number of serving and interfering satellites. Specifically, under the MA scheme with \( K = 96 \) and \( J = 72 \), the highest achievable rate is observed at an elevation angle of \( 10^\circ \), suggesting that the ground station can select a serving satellite with higher path gain while minimizing interference. However, the achievable rate decreases as the elevation angle increases beyond this point. This is because as the latitude of the ground station increases, more satellites become visible, leading to stronger interference and thereby reducing the achievable rate. Furthermore, increasing the number of satellites and orbits can enhance the achievable rate, as observed in previous figures. This improvement occurs because a larger number of satellites and orbits provide the ground station with more opportunities to select a better serving satellite using the proposed MA scheme. Nonetheless, the performance advantage of having more satellites and orbits diminishes at higher latitudes, since the increased number of visible satellites intensifies interference, which dominates the overall system performance. \par
\begin{figure}
	\includegraphics[width=0.45\textwidth]{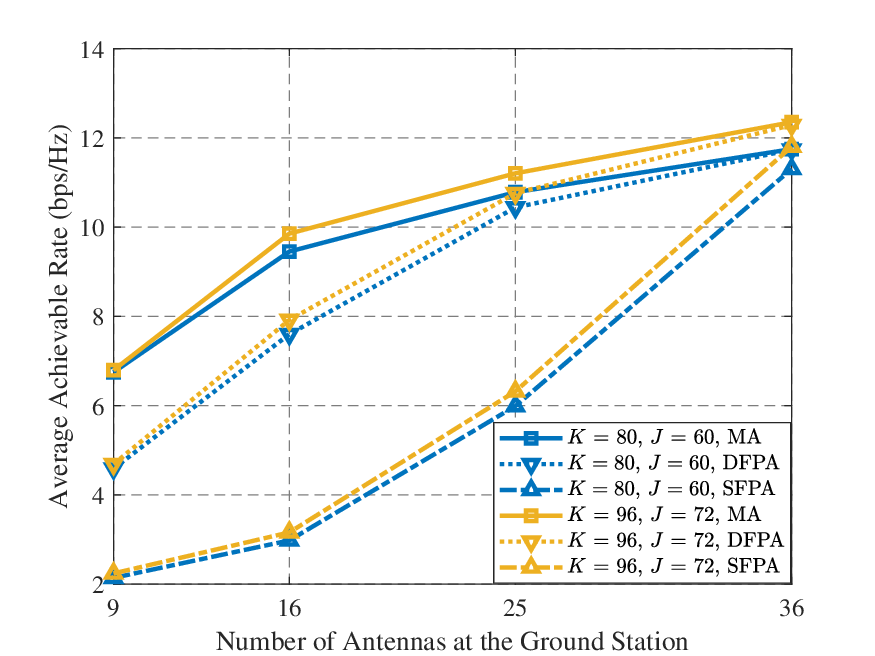}
	\caption{\footnotesize{The effect of the number of antennas at the ground station on the average achievable rate for different schemes and parameters.}}
	\label{figure5}
\end{figure}
Fig.~\ref{figure5} illustrates the impact of the number of receiving antennas at the ground station on the achievable rate performance, where the number of satellites per orbit is set to \( K = 80 \) and \( 96 \), and the number of orbits is configured as \( J = 60 \) and \( 72 \), respectively. It is observed that as the number of antennas increases, the average achievable rate improves across all schemes and parameter settings due to the higher degree of spatial freedom. Under the DFPA and SFPA schemes, where antenna positions are fixed, the achievable rate is lower than that of the MA scheme under the same parameters. Specifically, the proposed MA scheme further enhances rate performance by jointly optimizing antenna positions and beamforming. The MA scheme not only suppresses interference but also enables the ground station to connect to a better service satellite, thereby improving the achievable rate. Moreover, the rate improvement becomes more significant as the numbers of satellites and orbits increase. Comparing the MA schemes under different parameter settings shows that their performance gain becomes more pronounced with larger satellite and orbit numbers. For the DFPA and SFPA schemes, the performance gap between different configurations becomes more evident as the number of ground station antennas increases. Furthermore, the rate gain achieved by optimizing the antenna positions decreases as the number of antennas increases, as shown by the comparison between the MA and DFPA schemes with \( K = 96 \) and \( J = 72 \). This phenomenon occurs because the antenna movement capability is limited within the predefined antenna moving region \( \mathcal{A} \), which constrains the extent of achievable improvement in interference suppression and signal enhancement. \par
\begin{figure}
	\includegraphics[width=0.45\textwidth]{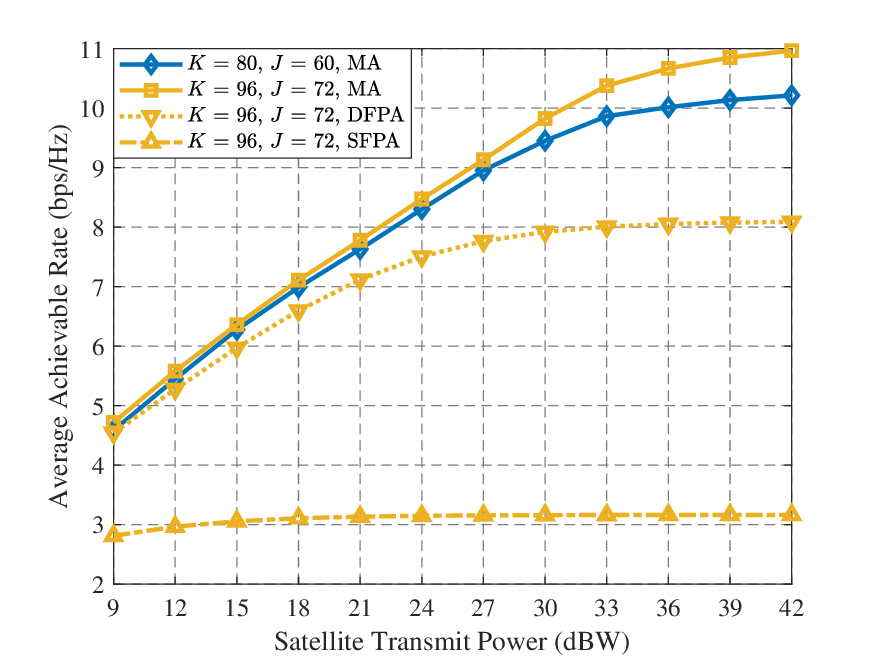}
	\caption{\footnotesize{The effect of satellite transmit power on the average achievable rate for different schemes and parameters.}}
	\label{figure6}
\end{figure}
Fig.~\ref{figure6} illustrates the effect of transmit power at each satellite on the average achievable rate, where \( K = 80 \) and \( 96 \), and \( J = 60 \) and \( 72 \), respectively. For the DFPA and SFPA schemes with fixed antenna positions and optimized beamformers, the average achievable rate increases gradually as the satellite transmit power increases. However, the rate gain from increasing transmit power diminishes at higher power levels. This is because higher transmit power introduces stronger interference, which eventually dominates and limits the rate gain achievable through beamforming optimization. Furthermore, the rate gain from increasing transmit power can be further amplified by implementing the proposed MA scheme, which jointly optimizes antenna positions and beamforming, especially under scenarios with a larger number of satellites and orbits. Moreover, even under the MA scheme, it is observed that the achievable rate does not always increase monotonically with higher transmit power, since other system parameters, such as the number of antennas, impose inherent performance limitations. Nonetheless, the rate gain achieved by combining the MA scheme and increased transmit power becomes more pronounced as the number of satellites and orbits increases, demonstrating that the proposed MA scheme is particularly effective in interference suppression for dense constellation scenarios. \par
\begin{figure}
	\includegraphics[width=0.45\textwidth]{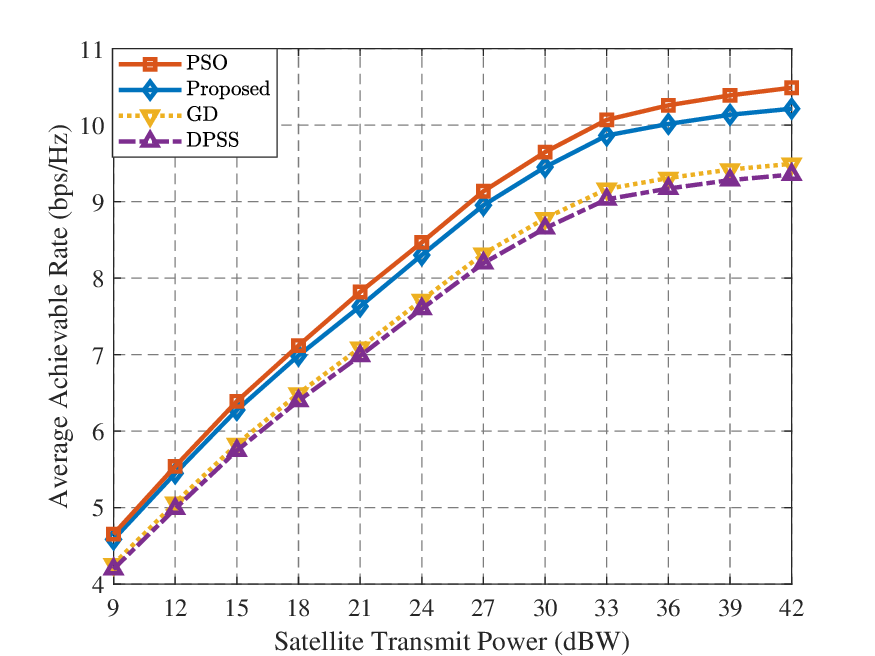}
	\caption{\footnotesize{The effect of satellite transmit power on the average achievable rate for different antenna position optimization schemes.}}
	\label{figure7}
\end{figure}
Furthermore, we compare our proposed SCA-based MA optimization algorithm with several alternative methods, including gradient descent (GD) \cite{10806489}, particle swarm optimization (PSO) \cite{10741192}, and discrete position sequential search (DPSS) \cite{10508218}. In these comparisons, the antenna positions are optimized using the respective algorithms, while beamforming vectors are still obtained by the proposed method for fairness. Fig.~\ref{figure7} illustrates the achievable average rate at the ground station with different MA optimization algorithms, where \( K = 80 \) and \( J = 60 \). The other parameters are the same as those in Fig.~\ref{figure6}. The results show that PSO achieves the best performance due to its global search capability, but it also incurs the highest computational complexity. The proposed SCA-based algorithm achieves performance close to PSO with significantly lower complexity, demonstrating its practical efficiency. GD and DPSS exhibit lower performance because of sensitivity to step size and limited search resolution, respectively. Overall, the proposed SCA-based MA optimization provides a good trade-off between increasing communication performance and decreasing computational complexity, which highlights its suitability for LEO satellite communication systems.
\section{Conclusions}
In this paper, we investigated achievable rate optimization for an MA-assisted LEO satellite communication network. We formulated a joint optimization problem to maximize the average achievable rate of the ground station by jointly optimizing its APV and AWVs under practical antenna movement and spacing constraints. To address this non-convex optimization problem, we applied the Lagrangian dual and quadratic transformations to derive a tractable formulation, and developed an efficient BCD-based iterative algorithm that alternately updates the APV and AWVs. Simulation results demonstrated that the proposed MA scheme significantly outperforms conventional FPA schemes by reshaping steering vectors adaptively via antenna positioning to more effectively mitigate interference and enhance signal reception. We also analyzed the impact of key system parameters, including the number of satellites, orbits, antennas, and transmit power. The results showed that although higher satellite density increases interference, the MA scheme effectively leverages spatial diversity and adaptive antenna positioning to improve performance. Notably, the proposed scheme allows the ground station to dynamically select more favorable service satellites over time, thereby enhancing long-term rate performance. Additionally, achievable rate gains do not always scale with increasing transmit power due to interference saturation in dense constellations. Overall, our findings revealed the strong potential of MA-assisted ground stations for interference suppression in ultra-dense LEO satellite networks. Looking ahead, future work could extend the proposed framework to incorporate imperfect CSI and more sophisticated interference management strategies, such as robust beamforming. These directions would further broaden the applicability and robustness of MA-assisted LEO communications.

\balance
\bibliographystyle{IEEEtran}
\bibliography{refabrv}

\end{document}